\documentclass[acmtog]{acmart}

\usepackage{booktabs} 
\usepackage{overpic}
\usepackage{subcaption}
\usepackage{bm}
\usepackage{array}
\usepackage{textcomp}
\usepackage{caption}
\usepackage{xspace}
\usepackage{multirow}
\usepackage{makecell}
\usepackage{url}
\usepackage{wrapfig}
\usepackage{dblfloatfix}

\renewcommand\footnotetextcopyrightpermission[1]{} 

\usepackage[ruled]{algorithm2e} 

\SetAlFnt{\small}
\SetAlCapFnt{\small}
\SetAlCapNameFnt{\small}
\SetAlCapHSkip{0pt}

\acmJournal{TOG}

\begin{document}


\title{BSDF Importance Baking: A Lightweight Neural Solution to Importance Sampling General Parametric BSDFs}

\author{Yaoyi Bai}
\affiliation{
 \institution{University of California, Santa Barbara}
 \city{Santa Barbara}
  \state{CA}
 \country{USA}
  \postcode{93106}
}
  \email{yaoyibai@cs.ucsb.edu}
  
  \author{Songyin Wu}
\affiliation{
 \institution{University of California, Santa Barbara}
 \city{Santa Barbara}
  \state{CA}
 \country{USA}
  \postcode{93106}
}
  \email{s_wu975@ucsb.edu}
  
  \author{Zheng Zeng}
\affiliation{
 \institution{University of California, Santa Barbara}
 \city{Santa Barbara}
  \state{CA}
 \country{USA}
  \postcode{93106}
}
  \email{zhengzeng@ucsb.edu}
  
    \author{Beibei Wang}
\affiliation{
 \institution{Nankai University}
 \city{Tianjin}
 \country{China}
}
  \email{beibei.wang@nankai.edu.cn}
  
  \author{Ling-Qi Yan}
\affiliation{
 \institution{University of California, Santa Barbara}
 \city{Santa Barbara}
  \state{CA}
 \country{USA}
  \postcode{93106}
}
  \email{lingqi@cs.ucsb.edu}

\begin{abstract}
Parametric Bidirectional Scattering Distribution Functions (BSDFs) are pervasively used because of their flexibility to represent a large variety of material appearances by simply tuning the parameters. While efficient evaluation of parametric BSDFs has been well-studied, high-quality importance sampling techniques for parametric BSDFs are still scarce. Existing sampling strategies either heavily rely on approximations, resulting in high variance, or solely perform sampling on a portion of the whole BSDF slice. Moreover, many of the sampling approaches are specifically paired with certain types of BSDFs. In this paper, we seek an efficient and general way for importance sampling parametric BSDFs. We notice that the nature of importance sampling is the mapping between a uniform distribution and the target distribution. Specifically, when BSDF parameters are given, the mapping that performs importance sampling on a BSDF slice can be simply recorded as a 2D image that we name as importance map. Following this observation, we accurately precompute the importance maps using a mathematical tool named optimal transport. Then we propose a lightweight neural network to efficiently compress the precomputed importance maps. In this way, we have brought parametric BSDF important sampling to the precomputation stage, avoiding heavy runtime computation. Since this process is similar to light baking where a set of images are precomputed, we name our method importance baking. Together with a BSDF evaluation network and a PDF (probability density function) query network, our method enables full multiple importance sampling (MIS) without any revision to the rendering pipeline. Our method essentially performs perfect importance sampling. Compared with previous methods, we demonstrate reduced noise levels on rendering results with a rich set of appearances, from multiple-bounce microfacet conductors with anisotropic roughness, to layered materials and Disney principled materials.

\end{abstract}

%
%




%
%






\def\LL{{\mathcal{L}}}
\def\bu{{\mathbf{u}}}
\def\bp{{\mathbf{p}}}
\def\bn{{\mathbf{n}}}
\def\bx{{\mathbf{x}}}
\def\by{{\mathbf{y}}}
\def\bo{{\bm{\omega}}}
\def\dd{{\,\mathrm{d}}}

\newcommand{\Reals}{\mathbb{R}}

\newcommand{\todo}[1]{[\textcolor{red}{\textbf{TODO:}} #1]}
\newcommand{\lingqi}[1]{\textcolor{blue}{Lingqi: #1}}
\definecolor{mypink}{RGB}{219, 48, 122}
\newcommand{\yaoyi}[1]{\textcolor{mypink}{Yaoyi: #1}}
\newcommand{\beibei}[1]{\textcolor{blue}{Beibei: #1}}
\definecolor{zhenggreen}{RGB}{63, 126, 49}
\newcommand{\zheng}[1]{\textcolor{zhenggreen}{Zheng: #1}}
\newcommand{\songyin}[1]{\textcolor{cyan}{Songyin: #1}}

\begin{teaserfigure}
  \centering
      \begin{overpic}[width = \columnwidth]{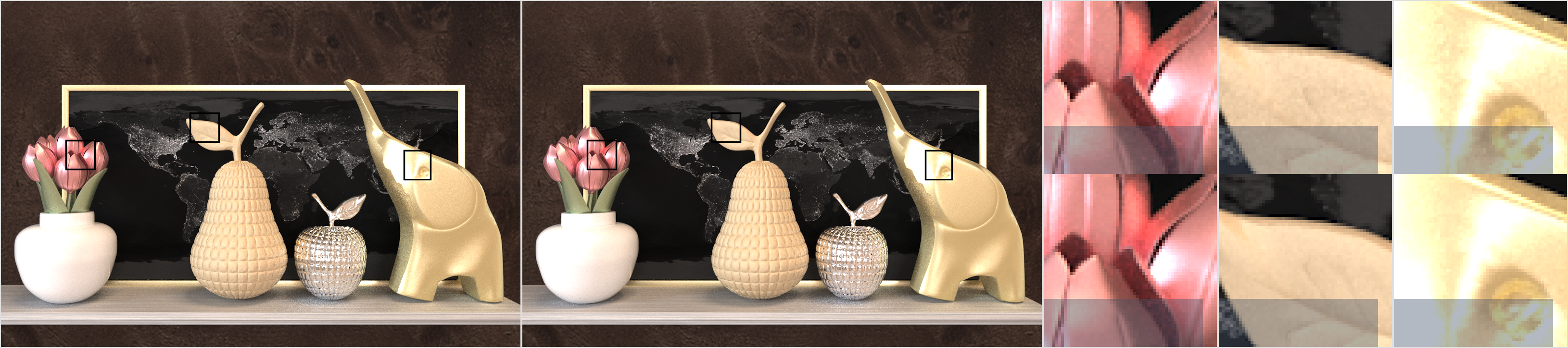}
    \put(1,0.5){\color{white}\small{\textbf{Heitz 256 spp (equal samples)}}}
    \put(34,0.5){\color{white}\small{\textbf{Ours 256 spp (equal samples)}}}
    \put(67,13){\color{white}\small{\textbf{Heitz et al.}}}
    \put(67,11.5){\color{white}\small{\textbf{relMSE $5.4e^{-3}$}}}
    \put(67,2){\color{white}\small{\textbf{Ours}}}
    \put(67,0.5){\color{white}\small{\textbf{relMSE $3.1e^{-3}$}}}
    \put(78,13){\color{white}\small{\textbf{Heitz et al.}}}
    \put(78,11.5){\color{white}\small{\textbf{relMSE $2.7e^{-3}$}}}
    \put(78,2){\color{white}\small{\textbf{Ours}}}
    \put(78,0.5){\color{white}\small{\textbf{relMSE $1.5e^{-3}$}}}
    \put(89,13){\color{white}\small{\textbf{Heitz et al.}}}
    \put(89,11.5){\color{white}\small{\textbf{relMSE $1.5e^{-2}$}}}
    \put(89,2){\color{white}\small{\textbf{Ours}}}
    \put(89,0.5){\color{white}\small{\textbf{relMSE $7.9e^{-3}$}}}
  \end{overpic}
\bigbreak
  \centering
         \begin{overpic}[width = \columnwidth]{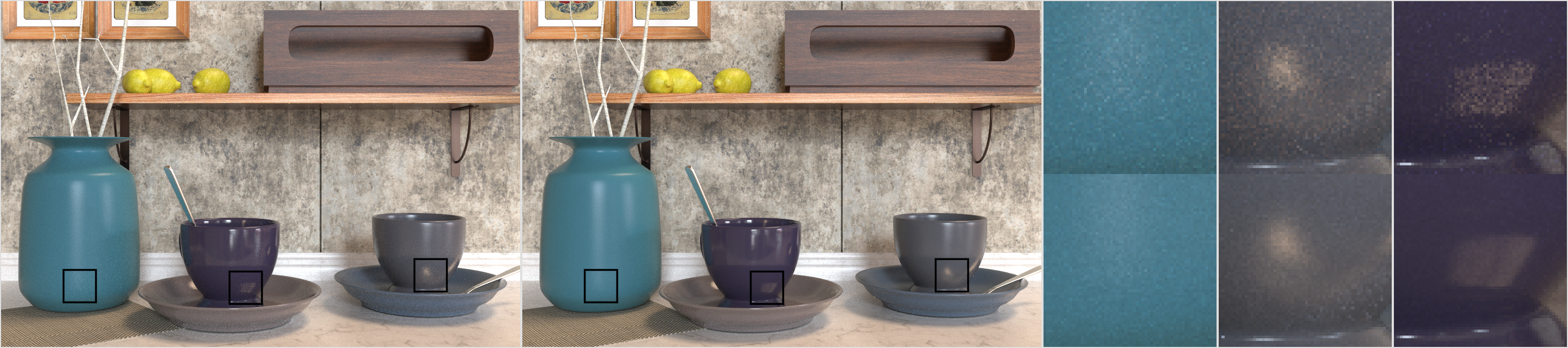}
  \put(1,0.5){\color{white}\small{\textbf{Guo et al. 850 spp (equal time)}}}
    \put(34,0.5){\color{white}\small{\textbf{Ours 1024 spp (equal time)}}}
    \put(67,12.5){\color{white}\small{\textbf{Guo et al.}}}
    \put(67,11){\color{white}\small{\textbf{relMSE $1.7e^{-3}$}}}
    \put(67,2){\color{white}\small{\textbf{Ours}}}
    \put(67,0.5){\color{white}\small{\textbf{relMSE $9.1e^{-4}$}}}
    \put(78,12.5){\color{white}\small{\textbf{Guo et al.}}}
    \put(78,11){\color{white}\small{\textbf{relMSE $1.2e^{-3}$}}}
    \put(78,2){\color{white}\small{\textbf{Ours}}}
    \put(78,0.5){\color{white}\small{\textbf{relMSE $5.3e^{-4}$}}}
    \put(89,12.5){\color{white}\small{\textbf{Guo et al.}}}
    \put(89,11){\color{white}\small{\textbf{relMSE $4.5e^{-4}$}}}
    \put(89,2){\color{white}\small{\textbf{Ours}}}
    \put(89,0.5){\color{white}\small{\textbf{relMSE $1.3e^{-4}$}}}
  \end{overpic}
  \vspace{-5mm}
    \caption{
    We demonstrate the effectiveness of our work (BSDF sampling network, together with BSDF evaluation and PDF query networks) on  \emph{multiple-bounce microfacet conductors} (top) and \emph{layered BSDF} (bottom). (Top) We compare our method with Heitz et al. ~\shortcite{heitz2016multiple} on various materials (roughness and colors) with equal sampling weight. Our method shows lower noise levels, despite requiring a longer time for network inference. (Bottom) Comparing with Guo et al. ~\shortcite{guo2018position} on several layered materials (different isotropic roughness, index of refraction, attenuation coefficient, and albedo colors) with equal time, our model achieves both higher performance and less noise level. }
    \label{teaser}
\end{teaserfigure}
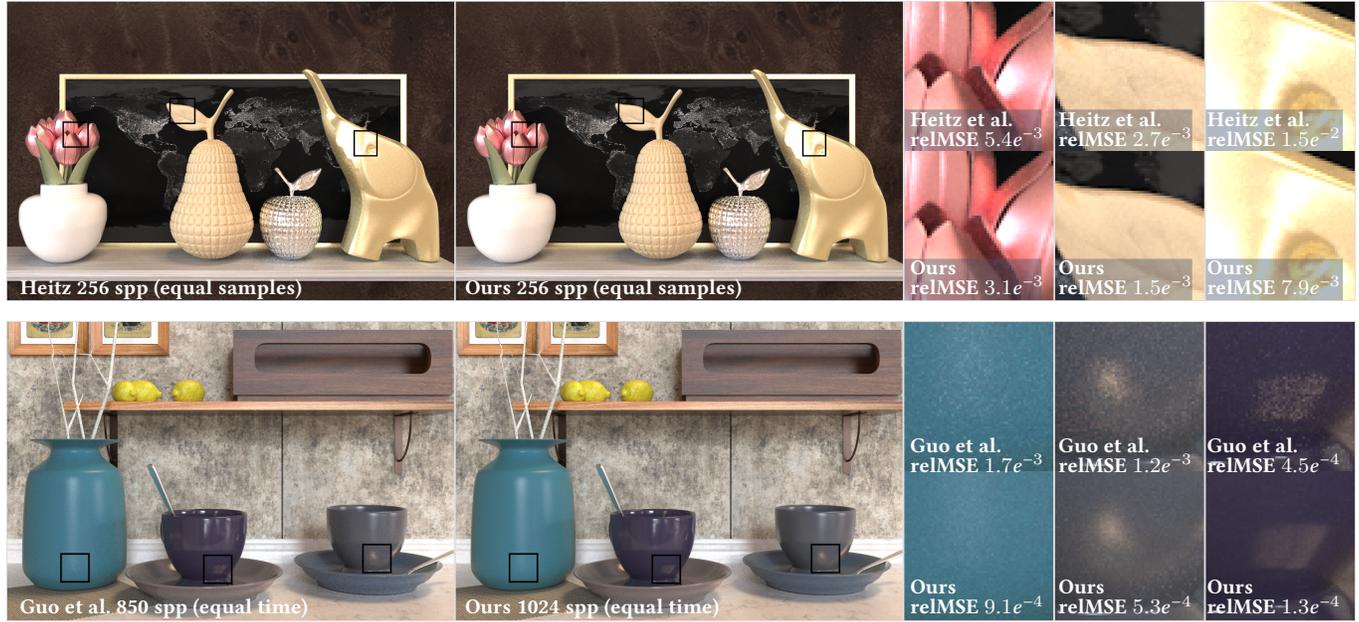

\maketitle
\section{Introduction}
\label{sec:intro}

Bidirectional Scattering Distribution Functions (BSDFs) are key to realistic appearances. Among all BSDFs, parametric BSDFs are pervasively used because of their flexibility to represent a large variety of materials' optical properties using a few parameters.

In the modern Monte Carlo rendering framework, BSDFs needs not only be evaluated, but also importance sampled. The core of BSDF importance sampling is to choose an incident direction according to a probability density function (PDF) that closely resembles the 2D BSDF slice jointly determined by BSDF parameters and an outgoing direction. However, parametric BSDFs barely have perfect importance sampling strategies. Some BSDFs do not support analytical sampling, e.g., the Disney Principled BRDFs using Generalized-Trowbridge-Reitz (GTR) ~\cite{burley2012physically, mcauley2012practical}. Some solutions tend to sample part of the BSDF slice, e.g., visible normal distribution functions (VNDFs) ~\cite{heitz2014importance} without taking the Fresnel term into account. Some other solutions use random walk to sample, e.g., position-free layered material \cite{guo2018position}, resulting in perfect importance sampling in theory but also high computational cost and variance.

Recently, neural approaches have been proposed to represent measured/synthetic BRDFs ~\cite{sztrajman2021neural,zheng2021compact,fan2021neural}, which are mostly tied with compression of BSDFs and seldomly study the problem of accurate importance sampling. Recent works that use neural methods for importance sampling do exist, but few work well with the naturally high-dimensional parametric BSDFs, especially those utilizing normalizing flow~\cite{Muller18,xie2019multiple}.

In this work, we focus on accurately solving the importance sampling problem for general parametric BSDFs. Our insight is that, importance sampling, regardless of the specific methods (e.g., marginalized inverse transform sampling) to achieve it, is in essence a mapping between a uniform distribution and a target distribution. Specifically, importance sampling a parametric BSDF slice is a 2D to 2D mapping which can be simply recorded as a 2D image that we name as \emph{importance map}. Suppose we can precompute and compress all the importance maps for all combinations of parameters and incident directions, we will be able to bring accurate parametric BSDF importance sampling to the precomputation stage, avoiding heavy runtime computation. 

Therefore, the core of general and accurate BSDF importance sampling is to generate high-quality importance maps. The commonly-used 2D importance sampling methods, e.g., the marginalized inverse transform sampling or hierarchical sample warping ~\cite{clarberg2005wavelet}, produce discontinuous importance maps, raising difficulties for both compression and interpolation. To overcome these issues, we introduce the Optimal Transport (OT) theory to this task, which is by nature suitable to provide a smooth mapping between distributions to generate importance maps efficiently. To alleviate the expensive storage cost of importance maps, we propose a lightweight neural network for compression and query. Together with a BSDF evaluation network and a PDF query network, we provide a complete neural solution for general parametric BSDFs, supporting full multiple importance sampling (MIS) without any revision to the rest of the rendering pipeline.

Since the process of computing and compressing importance maps is similar to light baking, in the sense that a set of images are precomputed and queried during runtime, we name our method \emph{BSDF importance baking}. It essentially performs perfect importance sampling, and can be used for any general parametric BSDFs, even those lacking analytic importance sampling solutions. Compared with previous methods, we demonstrate reduced noise levels on rendering results with a rich set of appearances, spanning a wide range of parametric BSDFs from multiple-bounce microfacet conductors with anisotropic roughness, to layered materials and Disney principled materials.

In summary, our contributions include:
\begin{itemize}
    \item a novel, accurate and efficient BSDF importance sampling solution for general parametric BSDFs,
    \item a new theory that connects optimal transport (OT) and BSDF importance sampling to compute \emph{importance maps}, which maps 2D uniform distribution to 2D BSDF slices.
    \item an application of lightweight neural networks to compress precomputed \emph{importance maps}, as well as optional BSDF evaluation and PDF query networks for comparatively complex BSDFs for a full MIS solution, and 
    \item a database of \emph{importance maps} that collects parametric BSDF importance sampling data for public use, including multiple-bounce microfacet conductors, layered BSDF, as well as Disney principled materials. 
\end{itemize}

\section{Related Work}
\label{sec:related}
\paragraph{Parametric BSDFs.} Parametric BSDFs represent different materials with explicit parameters. Two major groups of parametric BSDFs include empirical models (e.g.,\cite{phong1975, Ashikhmin2001}) and physically-based models, for example, microfacet models~\cite{blinn1977models} with different normal distribution functions (NDFs), including Beckmann \cite{BeckmannSpizzichino:1963}, GGX~\cite{walter2007microfacet} and Generalized-Trowbridge-Reitz (GTR)~\cite{burley2012physically}, and their multiple-bounce extensions~\cite{heitz2016multiple, Feng2018multiV, wang2022position}. Apart from multiple-bounce microfacet models, position-free layered BSDFs \cite{guo2018position} contain more parameters, such as the number of layers and the properties of the medium between layers. The parameters can also be artist-driven. For example, Disney principled materials ~\cite{burley2012physically, mcauley2012practical} are defined by few intuitive parameters, such as sheen and metallic. In this paper, We focus on providing a general solution to importance sampling parametric BSDFs. 


\paragraph{BSDF Importance Sampling.} Under the Monte Carlo rendering framework, the BSDF sampling strategy is crucial to variance reduction. However, sampling all the components within the BSDF is non-trivial. In the microfacet model, sampling the NDF gives a good approximation but can produce significant variance when the outgoing direction is from the grazing angle. Heitz et al.~\shortcite{heitz2014importance,heitz2018sampling, heitz2017simpler} reduced sampling variance by sampling the distribution of visible normals (VNDF). However, VNDF sampling only works for NDF that is stretch-invariant. Thus, some NDFs (e.g., GTR) do not allow for accurate VNDF sampling and only use an approximation. Even worse, when considering multiple scattering in the microfacet model~\cite{heitz2016multiple}, the importance sampling becomes a random walk due to the absence of the closed-form formulation, leading to low performance. {BSDF slices can also be approximated by simple lobes, such as two Gaussians~\cite{Fan:2022:NLBRDF} or Blinn-Phong models~\cite{sztrajman2021neural}, then BSDF importance sampling can be evaluated by sampling these known models. However, these simple lobes are unlikely to fit complex BSDFs and result in high variance, especially BSDFs with multiple lobes. However, our method computes the actual PDF values, producing less variance. \footnote{Note that BSDF is a 3-channel value, but PDF is a single-channel value. In this case, BSDF sampling cannot be perfect since PDF cannot always be the same with R, G and B values. Therefore, given some PDF values (for example, grayscale) and performing importance sampling accordingly, it should be considered the optimal solution for BSDF importance sampling. }  }

{\paragraph{Hierarchical Sample Warping.} Clarberg et al.~\shortcite{clarberg2005wavelet} proposed an efficient and high-quality hierarchical warping technique that maps a uniform distribution to a hierarchical distribution. They applied it to sample environment maps, BSDFs, and their products on-the-fly without evaluating the full integral. However, hierarchical structures break the continuity of distributions and increase the precomputation difficulty. Additionally, dynamic sampling requires that all BSDF slices are known and represented as wavelets. It means that all possible BSDFs need to be tabulated. However, tabulating all data is impossible for our layered BSDFs, which include ten dimensions. We will explain this issue in Sec.~\ref{sec:sampling_strategies}. }

{\paragraph{Sampling Specific Types of BSDFs} Lawrence et al.~\shortcite{lawrence2004efficient} reparameterize BRDFs and decompose them into factored representation to achieve simple and compact importance sampling of analytical and measured BRDFs. However, the decomposition structure can only handle simple BSDFs with mild glossiness or a single lobe. Otherwise, the results are far from accurate. Moreover, the parametric BSDFs investigated in this paper contain tremendously higher dimensions than 4D. Then decomposing these BSDFs becomes entirely not feasible.}


{Unlike these methods, our method considers 2D slices of the entire BSDF and allows perfect importance sampling. Our model does not require that the BSDF has a closed-form formulation or contains only one lobe, thus can be used for arbitrary parametric BSDF models, including multiple scattering models, layered BSDFs, and Disney principled BSDFs. }

\paragraph{Optimal transport in computer graphics.} Optimal transport (OT) is a mathematical framework to manipulate distributions~\cite{monge1781memoire, kantorovich1942translocation}. Recent works in computer graphics have applied optimal transport to various fields, such as shape interpolation~\cite{bonneel2011displacement,solomon2015convolutional,bonneel2016wasserstein}. OT is capable of providing good distribution mapping and natural interpolation, but the fact that OT calculates slowly constrains its computer graphics applications. Solutions with approximations enable faster OT calculation, such as Sinkhorn distances~\cite{cuturi2013sinkhorn}, convolutional Wasserstein distances~\cite{solomon2015convolutional},  geomloss~\cite{feydy2019interpolating}, sliced optimal transport (SOT)~\cite{paulin2020sliced} and sliced partial optimal transport (SPOT)~\cite{BC19}. However, they are still prohibitively slow in rendering, because computational resources have already been significantly diluted by the massive number of shading computations running in parallel. To avoid extensive runtime distribution mapping computation, we take advantage of the smooth mapping acquired using OT while entirely steering away from its disadvantage of slow computation by applying it only in the precomputation stage.

\paragraph{Neural network for BSDF representation.} Neural networks (NN) have been recently used for BRDF representation on measured materials by representing one spatially-varying BRDF (SVBRDF) or one Bidirectional Texture Function (BTF) per network~\cite{Rainer2019Neural,kuznetsov2021neumip}, all BRDFs within a unified network~\cite{Rainer2020Unified,hu2020deepbrdf, zheng2021compact}, or each BRDF as a standalone decoder network~\cite{sztrajman2021neural}. Besides compressing the measured BRDFs, Sztrajman et al.~\cite{sztrajman2021neural} also support importance sampling by mapping a measured BRDF to an approximate parametric BRDF, but the differences between them result in imperfect importance sampling. Zheng et al. \shortcite{zheng2021compact} applied two-layer NICE \cite{dinh2014nice} to sample measured BRDFs. Xie et al.~\shortcite{xie2019multiple} proposed to use a RealNVP \cite{dinh2016density} network to learn multiple scattering equivalent NDFs in the slope space, enabling importance sampling. However, their method cannot support the high-dimensional parameter space for parametric BSDFs (e.g., spatially-varying Fresnel and anisotropic roughness) because of the notoriously bulky structure of normalizing flow structures and is considered too slow~\cite{Muller18} for practical use.

Our method focuses on importance sampling general parametric BSDFs. By completely decoupling the computation of the importance maps to the precomputation stage, we only require using a lightweight neural network to perform compression. Furthermore, since our importance maps are generally smooth thanks to optimal transport, they are naturally suitable for a neural network to compress.

\begin{figure*}[t]
\begin{subfigure}{0.3\textwidth}
    \includegraphics[width = \columnwidth]{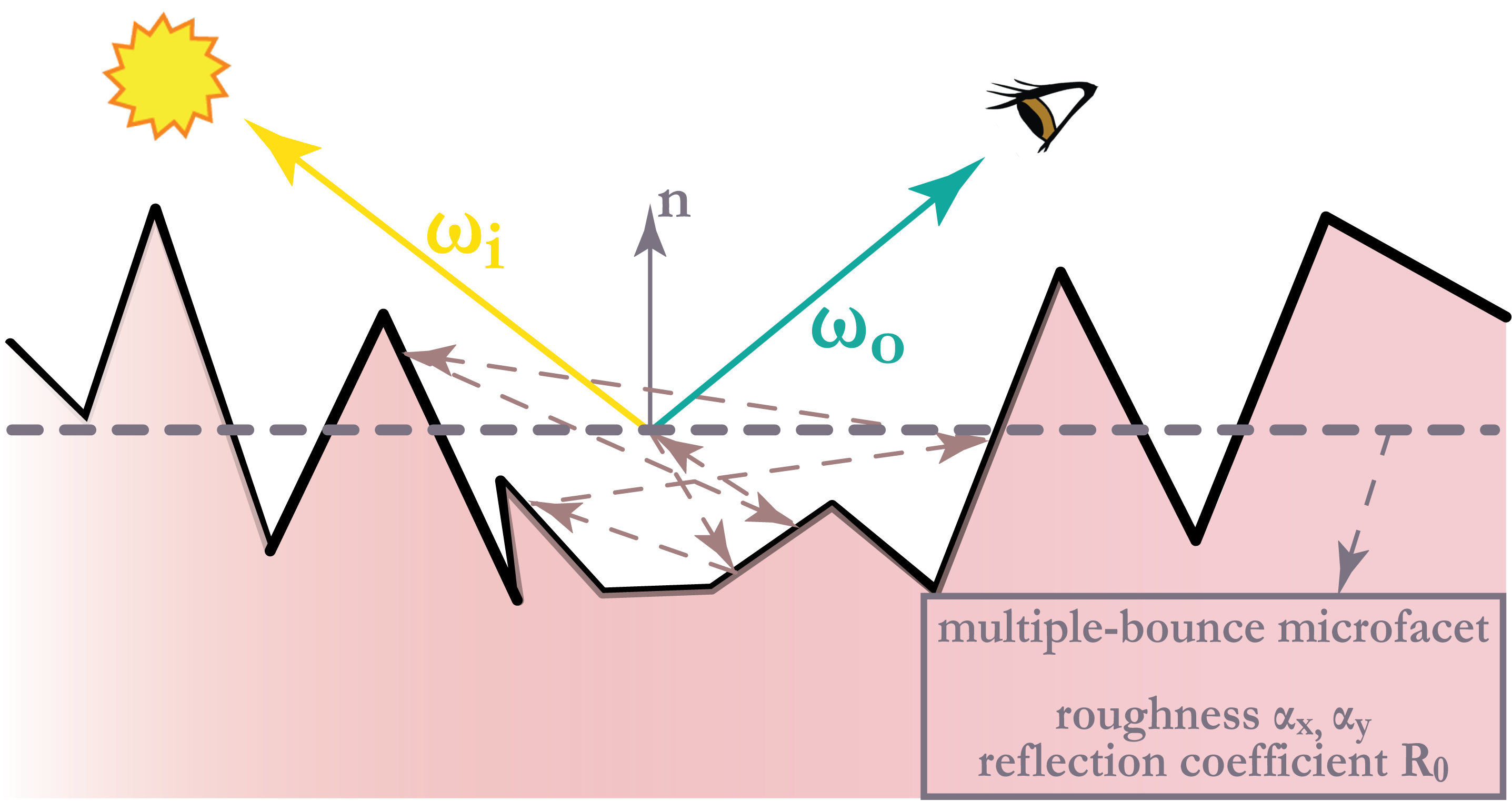}
    \caption{Multiple-bounce Microfacet BSDF}
\end{subfigure}
\begin{subfigure}{0.4\textwidth}
    \includegraphics[width = \columnwidth]{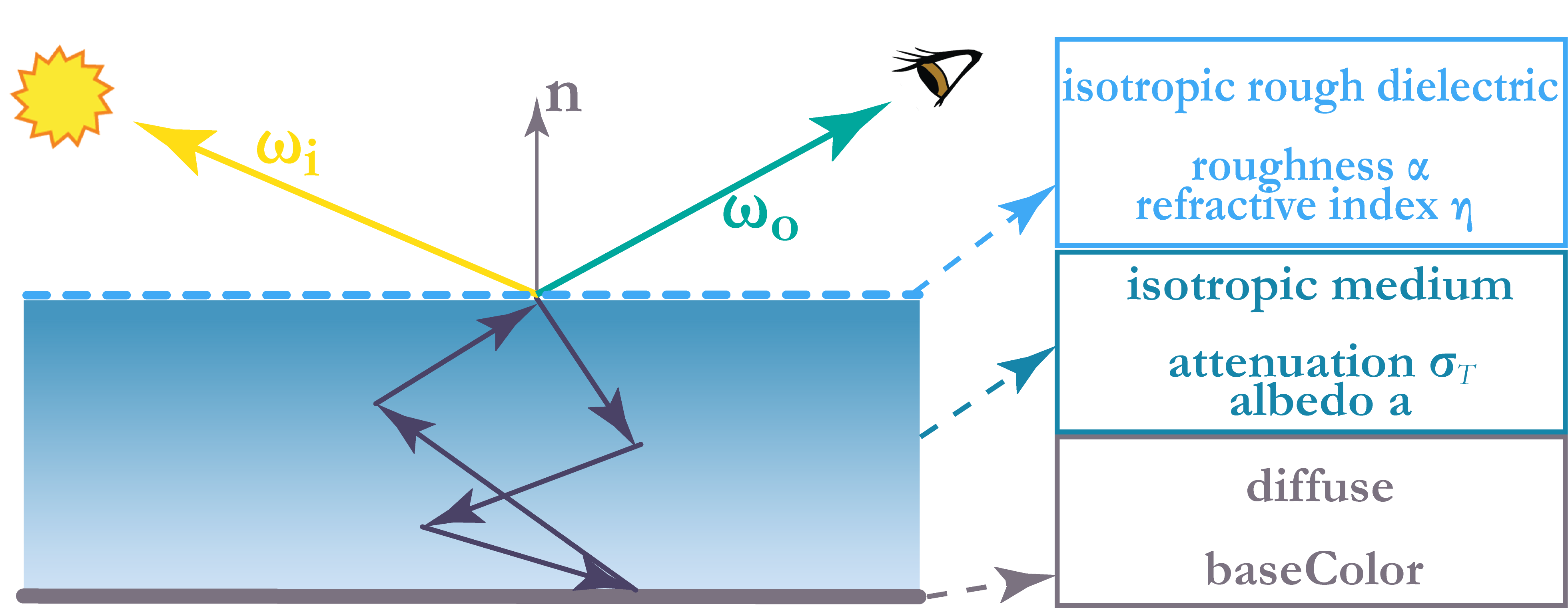}
    \caption{Two-layer Position-free Layered BSDF}
\end{subfigure}
\begin{subfigure}{0.25\textwidth}
    \includegraphics[width = \columnwidth]{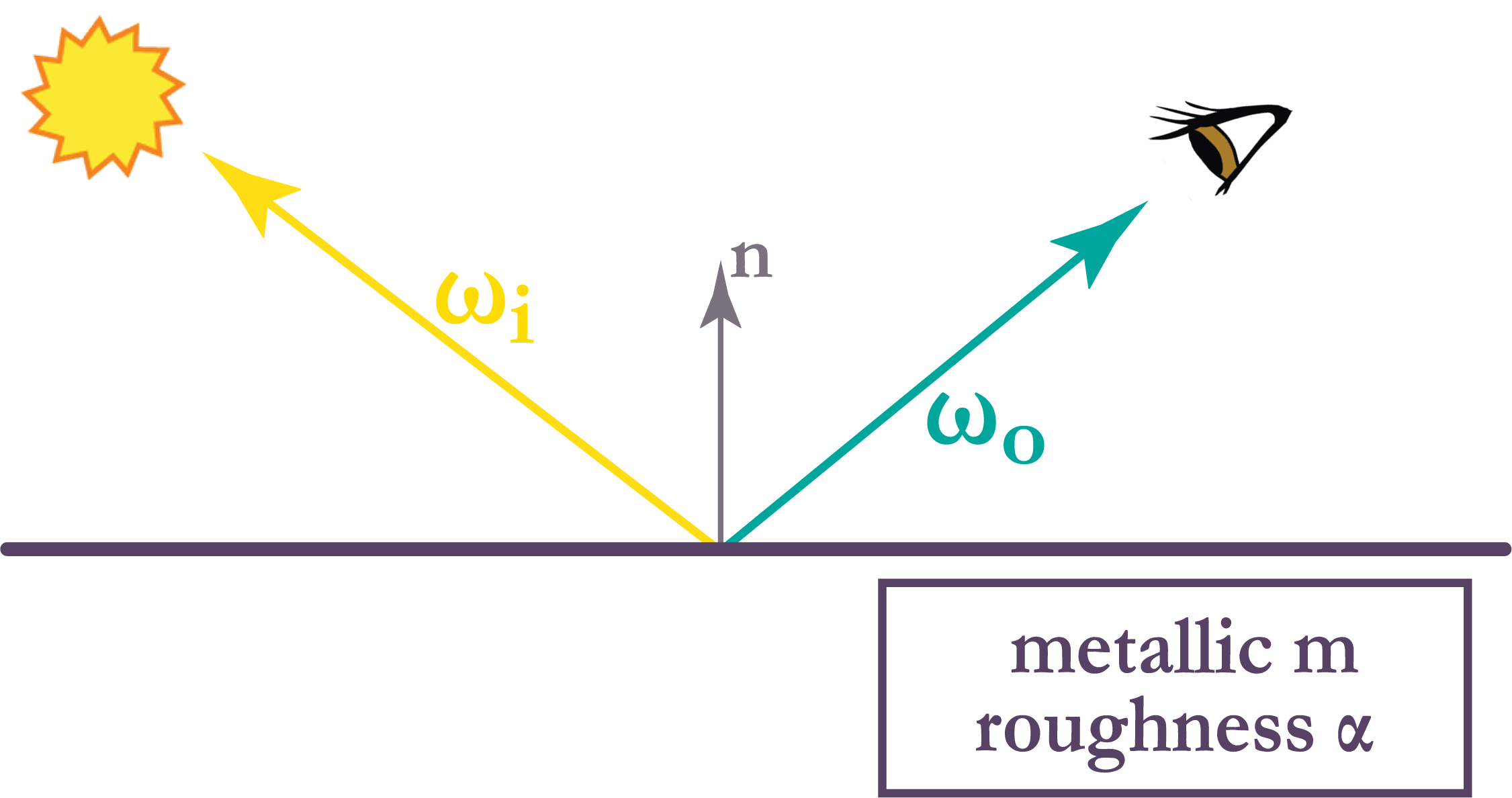}
    \caption{Disney Principled BRDF}
\end{subfigure}
    \caption{Several parametric BSDF models investigated in our paper. Note that the incident direction $\omega_{i}$ points towards light sources, and the outgoing direction $\omega_{o}$ points towards the camera.}
    
    \label{fig:microfacet_and_layered_bsdf}
\end{figure*}

\section{Background and Analysis}
\label{sec:Background_analysis}

\subsection{Parametric BSDF Rendering}
\label{sec:parametric_bsdf_rendering}

\paragraph{Parametric BSDFs}


Parametric BSDFs are explicitly controlled by material property parameters, such as roughness, refractive index, attenuation coefficient, and etc. For example, single-bounce Smith microfacet BRDFs can be analytically written as: 
\begin{equation}
    f_{s}(\bo_{i}, \bo_{o}) = \frac{F(\bo_{i})  G(\bo_{i}, \bo_{o}) D(\bo_{h})}{4 |\bo_{i}\cdot\bn| |\bo_{o}\cdot\bn|}, 
    \label{equ:single_microfacet}
\end{equation}
where $F$ is the Fresnel term, often approximated with a Schlick reflectance $R_0$ and an incident direction $\bo_{i}$. $G$ is the shadowing-masking term accounting for self-occlusion, and $D$ is the normal distribution function (NDF), parameterized on the half vector $\bo_{h}=\frac{\bo_{i}+\bo_{o}}{|\bo_{i}+\bo_{o}|}$, as well as roughness $\alpha_{x}$ and $\alpha_{y}$. Given these material parameters and an outgoing direction $\omega_{o}$, 2D BSDF slices can be easily generated where each pixel records the BSDF radiance value of an incident direction $\omega_{i}$.

For multiple-bounce Smith microfacet BSDFs where the light may bounce multiple times before exiting the surface as shown in Fig.~\ref{fig:microfacet_and_layered_bsdf}, there is still no analytical expression. Both the BSDF evaluation and importance sampling need heavy random walk simulations. In this work, we mainly focus on multiple-bounce BSDFs rather than single-bounce BSDFs, since the former are more challenging. Without loss of generality, we focus on GGX NDFs and describe conductors in the main text.


We also study layered BSDFs \cite{guo2018position}, as illustrated in Fig.~\ref{fig:microfacet_and_layered_bsdf}. The configuration for layered BRDFs includes a top
layer using a rough dielectric (with roughness $\alpha$ and refractive index $\eta$), a bottom layer using a diffuse BRDF and a homogeneous participating media (with attenuation coefficient $\sigma_{T}$ and albedo $(\mathbf{a}_{R}, \mathbf{a}_{G}, \mathbf{a}_{B})$) in the middle. Also note that there are no constraints about the number of layers or the types of BSDFs in each layer in our method. We merely initiate from a simple layered BSDF model to prove the generalization of our method.

Additionally, we study the Disney principled BSDF ~\cite{burley2012physically, mcauley2012practical}. This artists-friendly model was summarized to be few zero to one intuitive parameters, such as metallic, roughness, specular, anisotropic, sheen, and etc. In this paper, we mainly focus on manipulating metallic $\mathbf{m}$ and roughness $\alpha$ values also shown in Fig.~\ref{fig:microfacet_and_layered_bsdf}. 



\paragraph{Monte Carlo (MC) integration and importance sampling} The MC method provides an unbiased estimator to any definite integral in the domain $\Omega$ with an arbitrary integrand $f(x)$ 
\begin{equation}
  \int_{\Omega} f(x)\,\mathrm{d}x \approx \frac{1}{N} \displaystyle\sum\limits_{k=1}^n \frac{f(X_{k})}{p(X_{k})} \:\:\:\:\: X_{k} \sim p(x).
   \label{equ:MonteCarlo}
\end{equation}
The MC estimator approximates the integral  by drawing samples according to a probability density function (PDF) $p(x)$. The $f(X_k)/p(X_k)$ is known as the \emph{sampling weight}. To minimize the estimation variance, \emph{importance sampling} is desired -- the closer between the ``shapes'' of $f(x)$ and $p(x)$, the lower variance does the MC estimator have. 

\paragraph{BSDF importance sampling} MC integration is the core of the modern rendering pipeline, solving the \emph{rendering equation} at each shading point:
\begin{equation}
    L(\bo_{o}) = \int L(\bo_{i}) f_s(\bo_{i},\bo_{o}) \langle\bo_{i},\bn\rangle\,\mathrm{d}\bo_{i}.
\end{equation}
Given the BSDF parameters $\epsilon$ and the viewing direction $\bo_{o}$, the BSDF becomes a 2D slice, and BSDF importance sampling seeks on a good PDF similar to this 2D BSDF slice, together with a sampling technique that generates samples according to this PDF.

Though BSDF importance sampling can be easily calculated if the whole BSDF values at all dimensions are known, it would be extraordinarily difficult for parametric BSDFs because of storage constraints. At the same time, the efficiency of BSDF importance sampling is crucial, since it will be evaluated at every ray bounce for every sample ray from the camera. Massive amount of heavy BSDF importance sampling calculation will dramatically slow down the rendering calculation. Therefore, we propose to separate BSDF importance sampling calculation into the precomputation and compression of sampling data, as well as efficient runtime value lookup. 

\begin{figure}[t]
    \centering
    \includegraphics[width=\columnwidth]{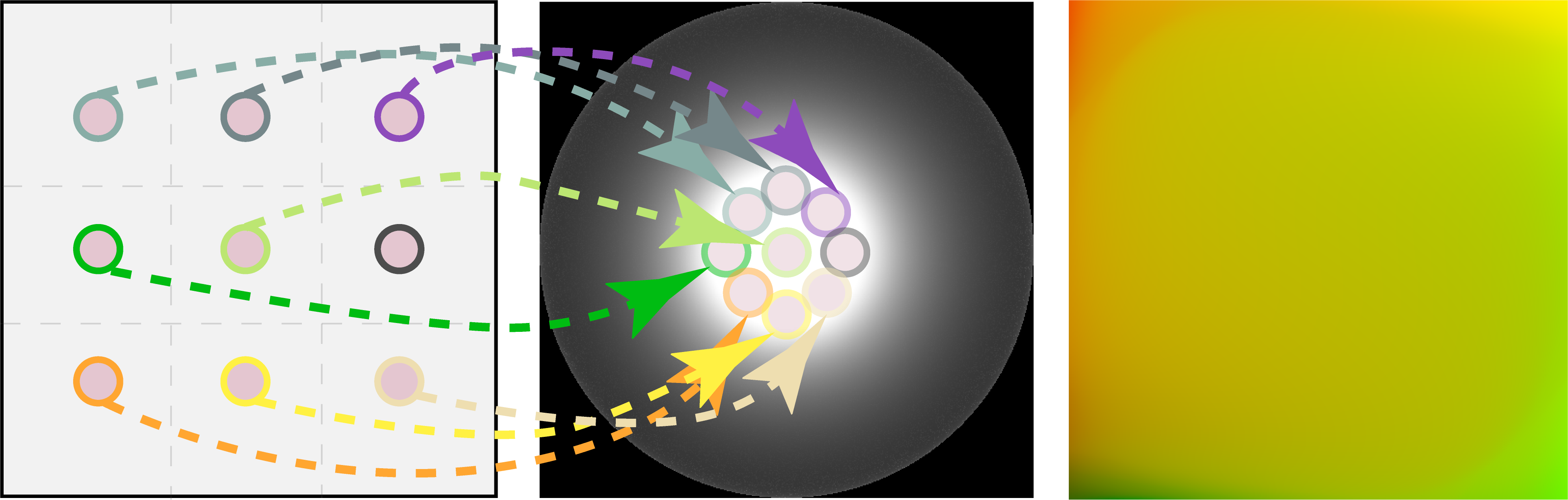}
\vspace{-6mm}
    \caption{A visualization of the process to generate an importance map. We start from generating sample points $(\xi_i^{(0)}, \xi_i^{(1)})$ on a uniform grid in the unit square $[0, 1]^{2}$ (left). The importance sampling process takes a 2D sample $(\xi_i^{(0)}, \xi_i^{(1)})$, and maps it to position $(u_i, v_i)$ on a 2D BSDF slice (middle). This mapping can be stored as an image (right), where the position of each pixel is $(\xi_i^{(0)}, \xi_i^{(1)})$, with its color set as $(r=u_i, g=v_i, b=0)$.    }
    \label{fig:sec3_importance_mapping}
\end{figure}

\subsection{BSDF Importance Baking}
\label{sec:Motivation_ImportanceBaking}

In this section, we analyze the essence of BSDF importance sampling, and present insight and motivation for our importance baking scheme.

We start with two issues from current BSDF importance sampling solutions.
\begin{enumerate}
    \item Not all parametric BSDFs can be analytically sampled, such as multiple-bounce microfacet BSDFs. This includes certain difficult-to-sample NDFs such as GTR, certain sampling methods such as visible NDF (VNDF) sampling only working with specific NDFs, as well as multiple-bounce BSDFs in general.
    \item Even analytic sampling methods are not perfect. In order to achieve the best quality, BSDF sampling requires a reasonable pdf $p(x)$ close to $f_{s} \langle\bo_{i},\bn\rangle$, but most analytical methods only sample the VNDF (ignoring the Fresnel term) or even just NDF $D$ (ignoring everything else). 
\end{enumerate}

 \begin{figure}[b]
    \centering
    \includegraphics[width=0.95\columnwidth]{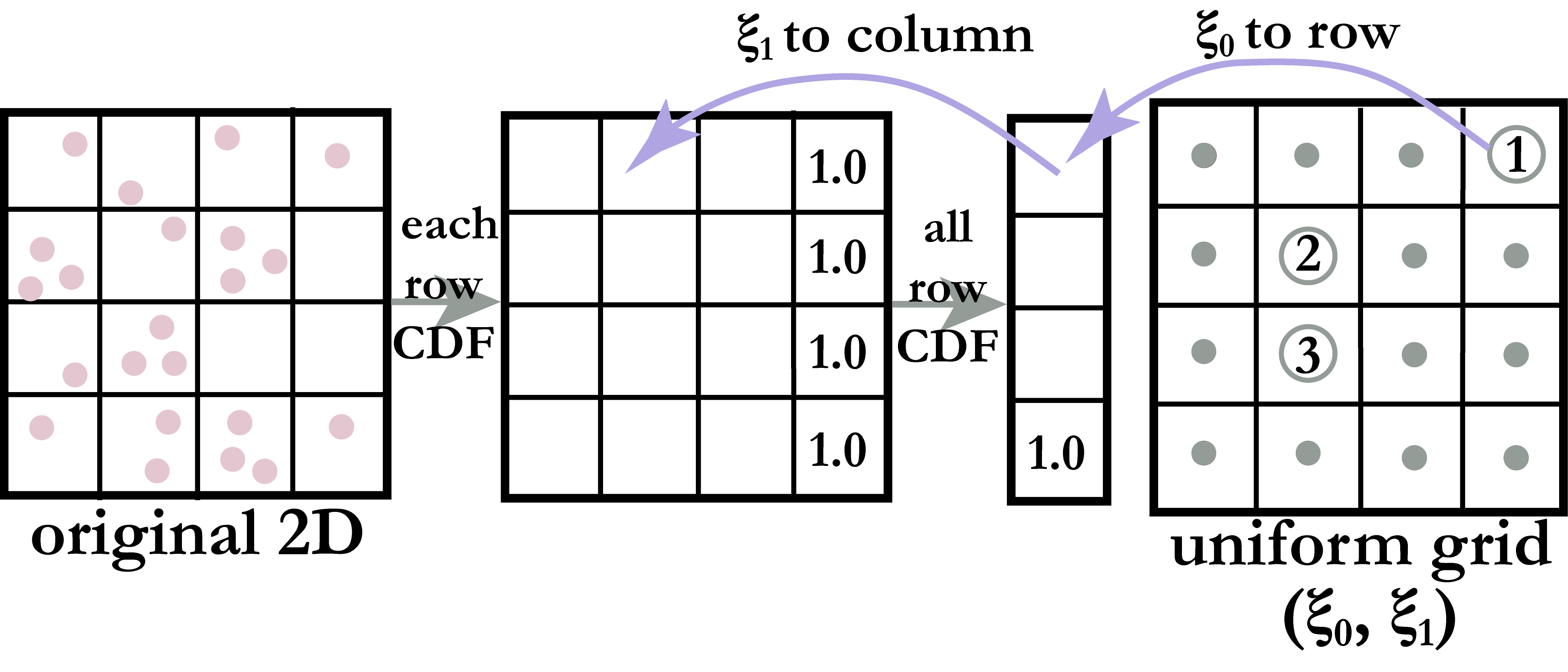}
    \caption{In the 2D marginalized inverse transform sampling, two random variables are used to firstly choose a row and then choose a column, forming a sampled location at the target distribution. Therefore, this sampling strategy acts as a mapping from the uniform distribution to the target distribution.}
    \label{fig:supply_inversecdf}
\end{figure}

To deal with these issues and achieve perfect importance sampling, we propose our understanding in the essence of the sampling process -- drawing samples according to a specific PDF is actually mapping from a uniform distribution to that PDF, viz., $f:(\xi^{(0)}, \xi^{(1)}) \mapsto (u, v)$ (in 2D). As an example, we visualize such a mapping acquired on a 2D BSDF slice in Fig.~\ref{fig:sec3_importance_mapping}. In this visualization, we assume the pixels represent a uniform grid on the unit square $[0,1]^2$, and each pixel $i$ at position $(\xi_i^{(0)}, \xi_i^{(1)})$ stores its mapped position $(u_{i}, v_{i})$ as red and green. In this way, we are able to generate an image for any 2D mapping, which we name the \emph{importance map}.

With this understanding, we immediately come up with the following insights.
\begin{enumerate}
    \item For a 2D BSDF slice, it is not crucial whether it can be sampled analytically or not because analytic sampling only corresponds to a quick lookup on the importance map. Instead, being able to acquire and query a high-quality importance map is the actual key to perfect importance sampling. Fortunately, we demonstrate that the importance map \emph{can be precomputed} with the help of optimal transport (Sec.~\ref{sec:OptimalTransport}).
    \item Suppose one importance map can be obtained from a 2D BSDF slice defined with BSDF parameters $\epsilon$ and the outgoing direction $\bo_{o}$. In order for the full parametric BSDF to be importance sampled, we have to collect all importance maps of all combinations of $\epsilon$ and $\bo_{o}$. This requires heavy storage, and therefore \emph{compression is needed}. Fortunately, we demonstrate that compression is not only possible but also efficient with the help of a lightweight neural network (Sec.~\ref{sec:NetworkForImportanceBaking}).
\end{enumerate}

Since we propose to precompute and compress the importance maps, the entire process is similar to the concept of light baking in real-time rendering. Therefore, we name it \emph{BSDF importance baking}.

\begin{figure}[t]
    \includegraphics[width = \columnwidth]{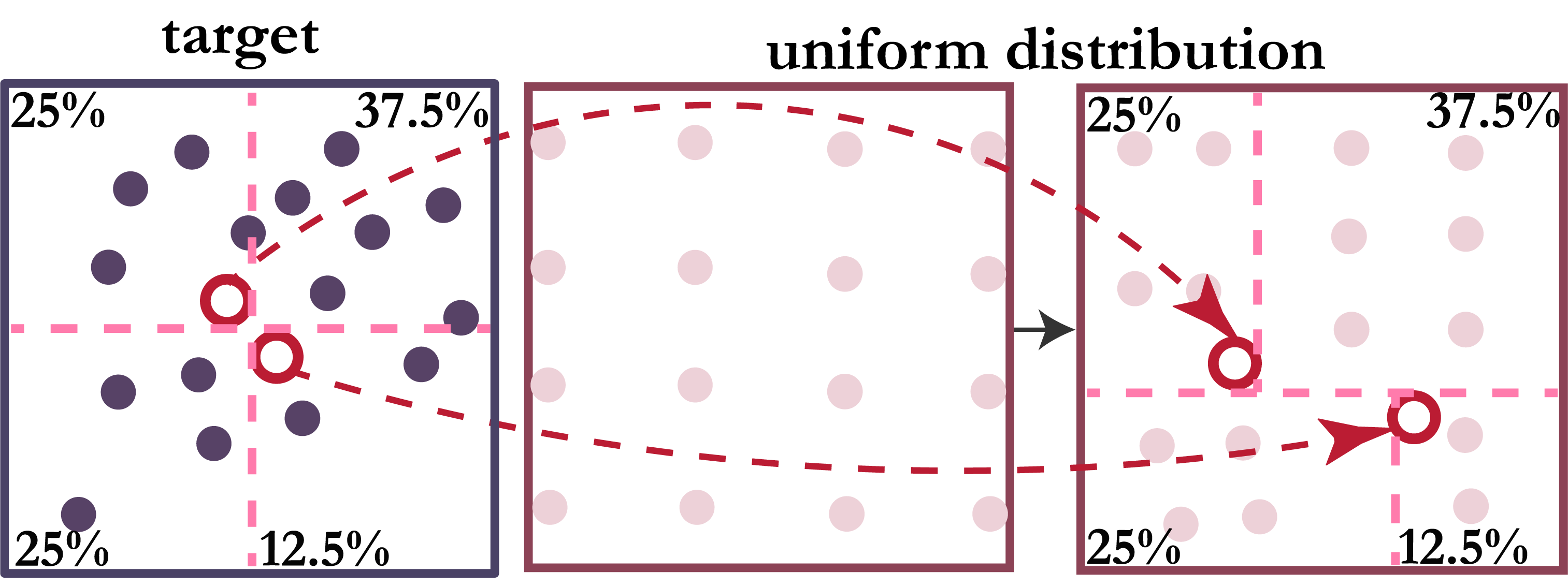}
    \caption{  The hierarchical sample warping strategy performs sampling by hierarchically choosing one of the quarters with its probability (i.e., the sample distribution in the target). However, this strategy can not guarantee continuity between each quarter. For example, two adjacent samples (red circles) in the target distribution are mapped from two samples at a distance, leading to the discontinuities in the binning result, as shown in Fig.~\ref{fig:comp_IS_wavelet_OT}.}
    \label{fig:wavelet}
\end{figure}

\begin{figure*}[t]
    \centering
    \begin{subfigure}{0.14\textwidth}
    \includegraphics[width=\columnwidth]{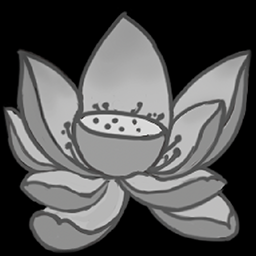}
    \caption{Input}
    \end{subfigure}
    \begin{subfigure}{0.28\textwidth}
    \includegraphics[width=\columnwidth]{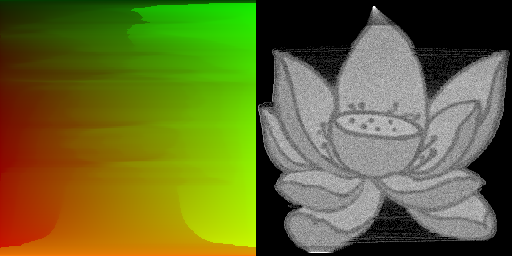}
    \caption{Marginalized Inverse Transform}
    \end{subfigure}
    \begin{subfigure}{0.28\textwidth}
    \includegraphics[width=\columnwidth]{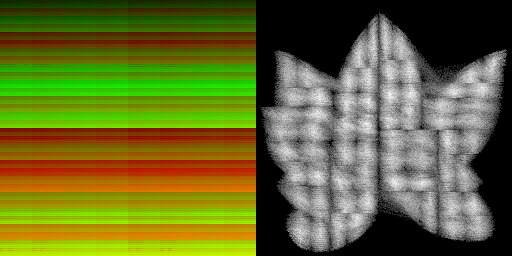}
    \caption{Hierarchical Sample Warping}
    \end{subfigure}
    \begin{subfigure}{0.28\textwidth}
   \includegraphics[width=\columnwidth]{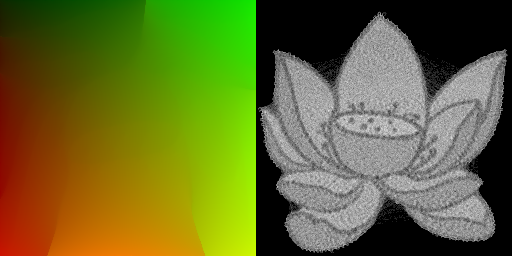}
    \caption{Optimal Transform}
    \end{subfigure}
\vspace{-3mm}
    \caption{ Let the input lotus image (a) be the target distribution. We use three importance sampling strategies to generate the importance map and then interpolate and reconstruct images from the importance map via binning. Here, \emph{binning} means sampling a location from the importance map using the same uniform sample sequence and accumulating it into the corresponding pixel, while applying bilinear interpolation, to reconstruct the images. Both marginalized inverse transform sampling and hierarchical sample warping generate importance maps with severe color discontinuity and high-frequency color variation, which leads to incorrect filamentous connections or grid artifacts in the binning result and raises difficulties for compression and further interpolation. On the contrary, the importance map by optimal transport has a smoother transition and lower frequency, allowing easier compression.}

    \label{fig:comp_IS_wavelet_OT}
\end{figure*}

\subsection{Existing Importance Sampling Strategies Analysis}
\label{sec:sampling_strategies}

Importance sampling estimates the properties of a distribution, which takes discrete values from a different function to reconstruct a new function that displays similarities with the target distribution. Usually from an easy-to-implement distribution, e.g., $U[0, 1]^d$ to an arbitrarily complex function. Importance sampling is not special, and it just requires that the target distribution is a normalized (constantly scaled) version of the target function.
 
In 1D, a typically used method is called the marginalized inverse transform sampling. It generates sample numbers using uniform random number $\xi \sim U[0, 1]$ between 0 and 1 for any probability distribution given its cumulative distribution function (CDF). Comparing $\xi$ and CDF values will map the random number to a certain value within the distribution domain. For discrete 2D situations, marginalized inverse transform sampling can be fulfilled by using two uniform random variables $(\xi_{0}, \xi_{1}) \sim U[0, 1]^{2}$ as shown in Fig.~\ref{fig:supply_inversecdf}. By calculating the CDF of all elements in one dimension (eg. all columns) and the overall CDF of the other dimension (eg. all rows), each pair of $(\xi_{0}, \xi_{1})$ can be used to attain a pair of 2D sample $(u_{i}, v_{i})$ on the target distribution similar to 1D situation. The mapping can be written as a function $f$ of elements between distributions $f:(\xi_{0}, \xi_{1}) \mapsto (u_{i}, v_{i})$. 

Another typical sampling strategy is hierarchical sample warping~\cite{clarberg2005wavelet}, which constructs levels of sample point mapping from coarse to detail between uniform distributed samples and target distribution, as shown in  Fig.~\ref{fig:wavelet}. It efficiently generates high-quality sample mapping without evaluating the whole target distribution. However, we observe that hierarchical sample warping allows two neighboring samples on the target to be mapped from two far-away samples in the uniform distribution (see Fig.~\ref{fig:wavelet}).

In Fig.~\ref{fig:comp_IS_wavelet_OT}, we compare different sampling strategies by showing their \emph{importance maps}, as well as the \emph{sampling binning results} by applying bilinear interpolation using each importance map. Both marginalized inverse transform sampling and hierarchical sample warping show discontinuous importance maps, leading to an incorrect filamentous connections or grid artifacts in the binning results. 

In summary, the existing solutions can be exploited for importance sampling, but they generate discontinuous importance maps, which raises difficulties for precomputation, compression, and interpolation. Therefore, we seek for a better solution to generate importance maps.




\section{Our method}

\subsection{Optimal Transport for Precomputed Importance}
\label{sec:OptimalTransport}

In this subsection, we focus on attaining the importance maps. We start from an important fact that is often ignored. That is, potential mappings that produce \emph{the same importance sampled PDF} are not unique. 
Consider a toy example of sampling a truncated, and normalized 1D Gaussian defined on $[0,1]$ shown on the right. Suppose we uniformly subdivide $[0,1]$ into four segments A, B, C and D, each integrates to a probability of $0.1$, $0.4$, $0.4$ and $0.1$. Then we subdivide the uniform $[0,1]$ into four 

\begin{wrapfigure}{r}{0.4\columnwidth} 
\hspace{-3mm}
    \includegraphics[width=0.4\columnwidth]{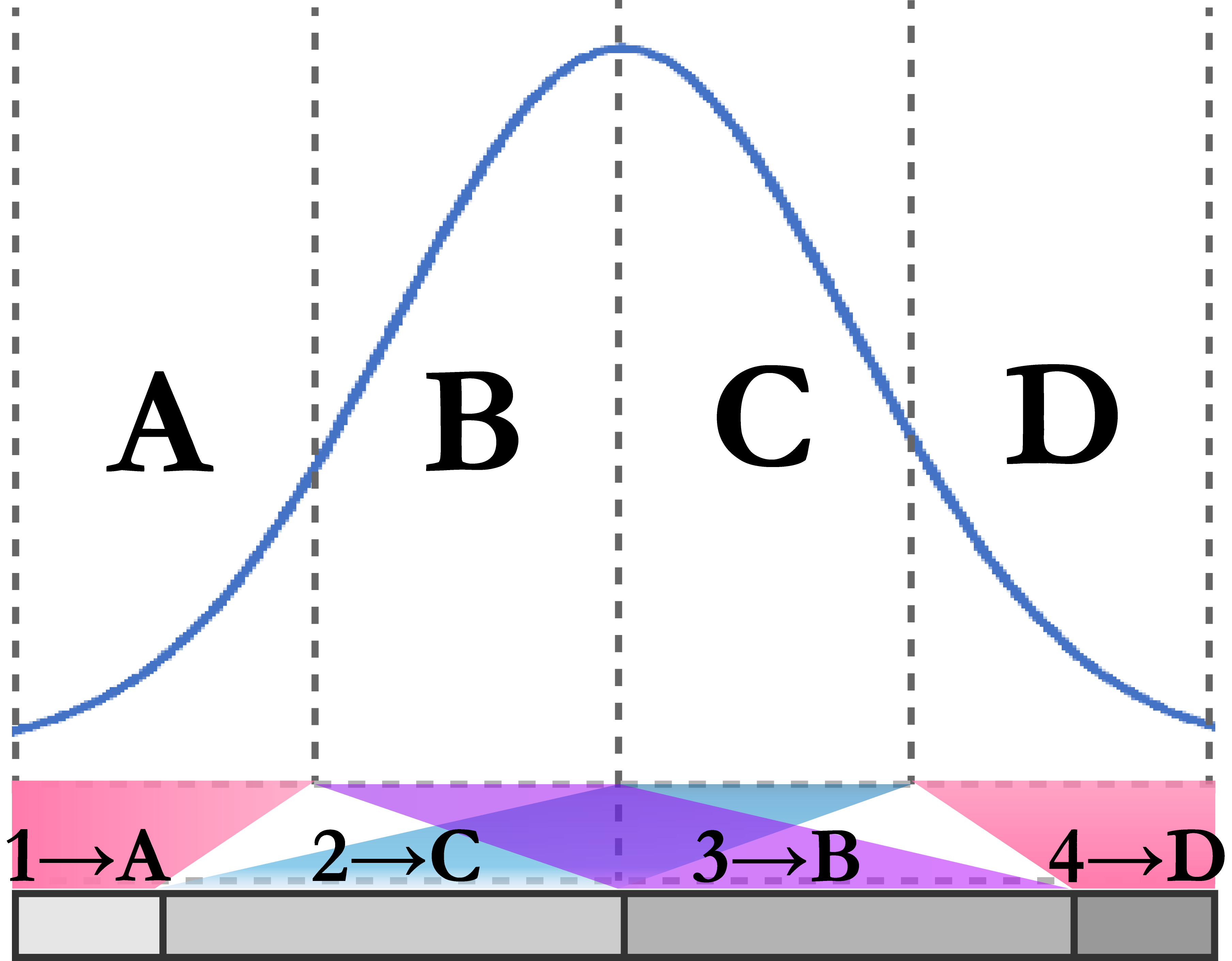}
    \label{fig:inline}
\vspace{-3mm}
\end{wrapfigure}

segments with lengths $0.1$, $0.4$, $0.4$ and $0.1$, and name them 1, 2, 3 and 4, respectively. Then the mapping $1\mapsto A$, $2\mapsto B$, $3\mapsto C$, $4\mapsto D$ is a valid mapping that importance samples the Gaussian, but the mapping $1\mapsto A$, $2\mapsto C$ and $3\mapsto B$, $4\mapsto D$ is also perfect importance sampling.

Moreover, from this example, one can immediately tell which sampling strategy is better: the first one is much smoother, and the second one suffers from discontinuity. This conclusion is never trivial because it directly proves that one pervasively used solution to obtain the importance maps -- the marginalized inverse transform sampling, a.k.a. row-column sampling -- is not suitable for generating good importance maps (Fig.~\ref{fig:comp_IS_wavelet_OT}). This is because a small perturbation in $\xi_0$ may result in a different row with a different 1D distribution, then even for similar $\xi_1$, the resulting column can be far away. 

Therefore, we prefer a strategy that provides smooth mapping that a small perturbation makes the mapped sample move moderately nearby. To satisfy this requirement, we refer to optimal transport (OT) -- specifically, discrete optimal transport from the Lagrangian view~\cite{feydy2019interpolating}, which is able to find an optimal one-to-one mapping between two \emph{point distributions} with the same number of points. In our case, this is to map from the unit square to the 2D BSDF slice for each combination of BSDF parameters $\epsilon$ and outgoing directions $\bo_{o}$.

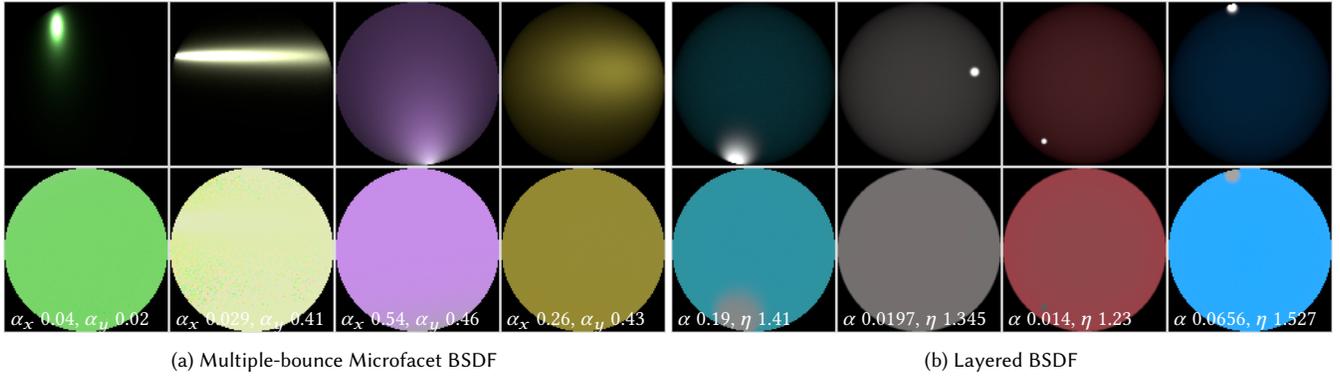
\begin{figure*}[t]
    \centering
    \begin{subfigure}{0.49\textwidth}
    \begin{overpic}[width=\columnwidth]{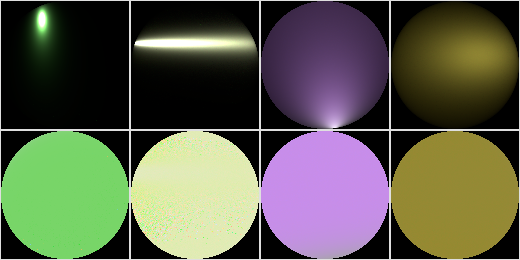}
    \put(1,1){\color{white}\small{$\alpha_{x}$ 0.04, $\alpha_{y}$ 0.02}}
    \put(26,1){\color{white}\small{$\alpha_{x}$ 0.029, $\alpha_{y}$ 0.41}}
    \put(51,1){\color{white}\small{$\alpha_{x}$ 0.54, $\alpha_{y}$ 0.46}}
    \put(76,1){\color{white}\small{$\alpha_{x}$ 0.26, $\alpha_{y}$ 0.43}}
    \end{overpic}
    \caption{Multiple-bounce Microfacet BSDF}
    \end{subfigure}
    \begin{subfigure}{0.49\textwidth}
    \begin{overpic}[width=\columnwidth]{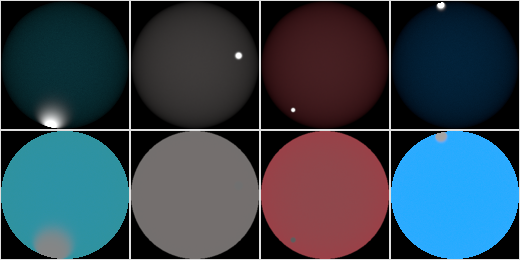}
    \put(1,1){\color{white}\small{$\alpha$ 0.19, $\eta$ 1.41}}
    \put(26,1){\color{white}\small{$\alpha$ 0.0197, $\eta$ 1.345}}
    \put(51,1){\color{white}\small{$\alpha$ 0.014, $\eta$ 1.23}}
    \put(76,1){\color{white}\small{$\alpha$ 0.0656, $\eta$ 1.527}}    
    \end{overpic}
    \caption{Layered BSDF}
    \end{subfigure}
    \vspace{-3mm}
    \caption{We visualize BSDF slices (top row) and their corresponding sampling weight slices (bottom row). (a) For multiple-bounce microfacet BSDFs, no matter what the BSDF looks like, the sampling weight is a low-frequency function whether the roughness values are low or high, or the BSDF is isotropic or anisotropic. Therefore, we can output sampling weight in the BSDF sampling network without increasing its complexity. Note that the noise of the sampling weight comes from the original BSDF slice. (b) For layered materials, the situation becomes more complex. In the two-layer model we are studying, the reflection on the top dielectric surface creates a white highlight circle. Therefore, the sampling weight images of layered BSDF are not in uniform color. However, the sampling weight slices at the bottom row still show that the sampling weights are relatively low-frequency values.  }
    \label{fig:sample_weight}
\end{figure*}

To conduct optimal transport, we first discretize both distributions into ordered point sets 
\begin{equation}
\bm{\alpha} = \cup_{i=1}^n \delta\left(\xi_i^{(0)},\xi_i^{(1)}\right),\quad \bm{\beta} = \cup_{j=1}^n \delta\left(u_j,v_j\right), 
\end{equation}
where $n$ is the total number of points, and the $\delta(\cdot)$ is Dirac delta impulse at different positions. We weigh each point the same, which immediately indicates that it is the local densities of those points that represent the values of the original continuous distributions. In other words, the continuous-to-discrete conversion itself is exactly importance sampling. We manually convert the unit square into a regular grid $\bm{\alpha}$ (pixels) and use row-column sampling to convert the 2D BSDF slice into $\bm{\beta}$.

Then we conduct optimal transport, giving a 1-to-1 correspondence between any two point distributions $\bm{\alpha}$ and $\bm{\beta}$, minimizing the Euclidean distance between them:
\begin{equation}
  \mathrm{argmin}_\psi \sum_{i=1}^n \Vert\bm{\alpha}(i)- \bm{\beta}(\psi(i))\Vert, 
\end{equation}
where $\psi$ is a permutation of the sequence $1,2,\dots,n$, computed by the optimal transport process. After this, we record the positions of each $\bm{\beta}(\psi(i))$ into each pixel's red and green channel, which completes the computation of an importance map.

Note that during the conversion of $\bm{\beta}$, we used row-column sampling. However, this is in essence different from using that to find the mapping -- we only use row-column sampling to discretize an image into points, and the mapping is found by optimal transport. More conversion tools can be explored in the research area of image stippling~\cite{kim2008feature}.

Also, note specifically that we focus on the \emph{black box} usage of optimal transport as a general mathematical tool. However, we do not intend to compare or improve specific optimal transport solvers. We also do not extend further discussion on specific accurate/approximate distance metrics (e.g., earth mover's distance, Wasserstein distance, Kullback–Leibler divergence, Sinkhorn distance, etc.). In Sec.~\ref{sec:suppl_datagen}, we provide our choices for implementation. 

In summary, for each of the three kinds of BSDFs and each possible combinations of parameters, we compute the BSDF slices, and calculate OT mapping to attain importance maps. With all these calculations, we have a database of importance maps and BSDF slices. Details of the whole generation process can be found in Sec.~\ref{sec:suppl_datagen}.

\subsection{Lightweight Neural Networks for Importance Baking}
\label{sec:NetworkForImportanceBaking}
Now that the OT already provides a reliable sampling scheme with these computed \emph{importance maps}, then we should focus on data storage and compression. We introduce a lightweight neural network for compression and storage because networks provide a higher compression rate and better expressiveness than traditional methods.

Before we proceed, we would like to note, as one could already tell immediately, the difference between our method and the normalizing flow methods. We treat the neural network as a general compression tool to save the data we have already computed for importance sampling. This is in essence different to letting neural networks figure out how to perform importance sampling. Therefore, our solution makes it much easier for networks to ``learn'' the sample mapping, and dramatically reduces the complexity of our networks, resulting in significantly better results and performance. We will also discuss more on the differences between our method and normalizing flow in Sec.~\ref{sec:properties_analysis}.



\begin{figure}[b]
    \centering
    \includegraphics[width=\columnwidth]{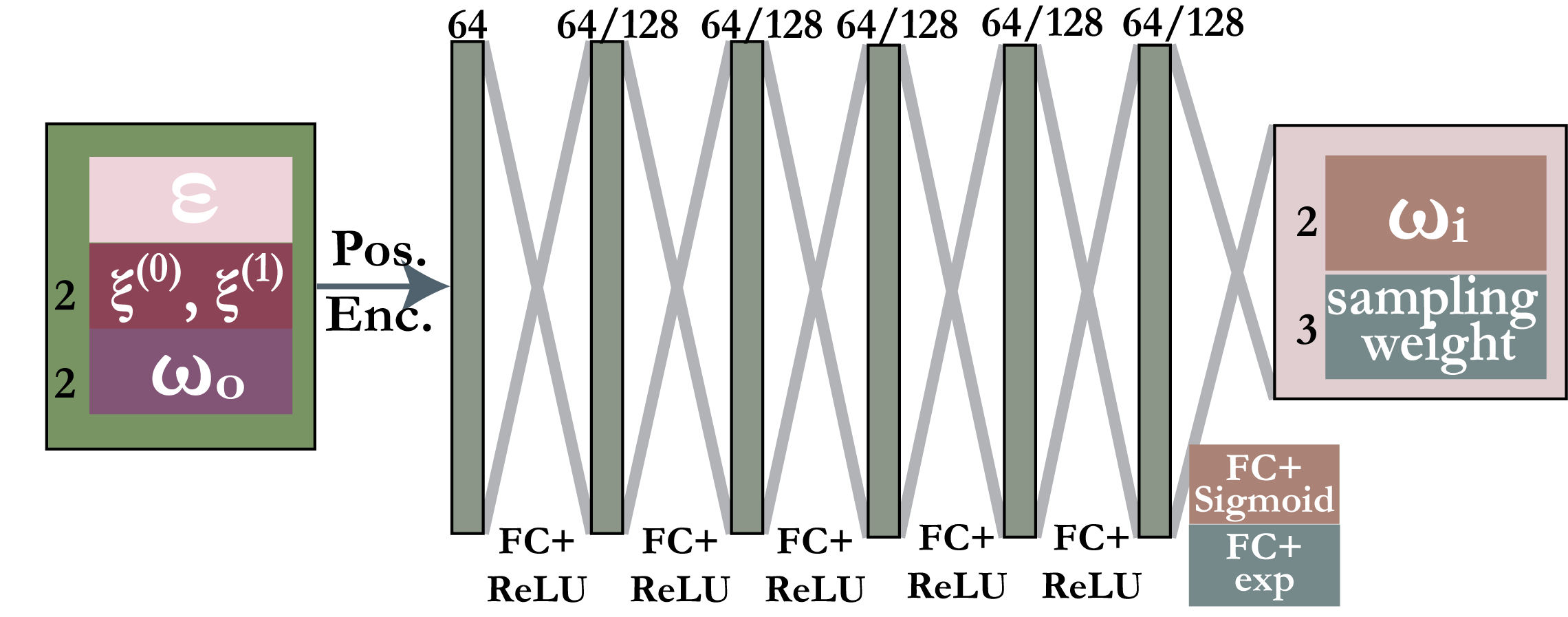}
    \vspace{-6mm}
    \caption{The structure of our BSDF importance sampling network. The network takes the following as inputs: material parameters $\epsilon$, outgoing ray direction $\omega_{o}$ and a pair of uniformly distributed random numbers $\xi_{0}$ and $\xi_{1}$. Then we process the inputs with a positional encoding, feed to an MLP (six layers of 64/128 hidden units), and finally output the sampled direction $\omega_i$ and the sampling weight. Note that we increase the MLP size from 64 to 128 for layered BSDFs, since the sampling weight slices are more complex than other BSDFs as shown in Fig.~\ref{fig:sample_weight}. More details are shown in Sec.\ref{sec:implemt_training}. }
    \label{fig:nn_sample}
    \vspace{-5mm}
\end{figure}

We start with a few design principles from a series of observations.
\begin{enumerate}
    \item We notice a significant amount of similarities between the importance maps when the BSDF parameters and incident directions change. The smooth change of these importance maps inspire us to use a neural network to compress them.
    \item As mentioned in Sec.~\ref{sec:related}, since runtime performance is crucial to core rendering, the highest level design of our neural network is to keep it as lightweight as possible, thus allowing for fast inference during rendering. 
    \item During BSDF importance sampling, only one incident direction needs to be sampled at a time. Therefore, the importance map should be \emph{point queried} instead of being output on the whole. This also further reduces the complexity of our neural network.
    \item For Monte Carlo estimation, the BSDF sampling process is expected to output not only an incident direction but also its \emph{sampling weight}, which is the BSDF value divided by the PDF value. For multiple-bounce microfacet materials and Disney principled materials, the sampling weight is a 3-channel value and is close to constant since we design our PDFs to have the same shapes with the BSDF slices converted to grayscale. For layered materials, the sampling weight data are more complex but still relatively low-frequency. Fig.~\ref{fig:sample_weight} shows more details about sampling weight data.

\end{enumerate}

Based on these design principles, we propose a lightweight neural network for our BSDF importance sampling.

\paragraph{BSDF Sampling Network}

Aside from the BSDF parameters $\epsilon$ and the outgoing direction $\bo_{o}$, our importance sampling network takes two random numbers and outputs the sampled incident direction together with its sampling weight:
\begin{equation}
   \mathcal{I}(\epsilon, \bo_{o}, \xi_{0}, \xi_{1}) = (\bo_{i}, \text{sw}_{R}, \text{sw}_{G}, \text{sw}_{B}), 
\label{equ:sample}
\end{equation}
where $\text{sw}$ is the 3-channel sampling weight. The network structure is shown in Fig.~\ref{fig:nn_sample}. Note that since we have full control of the sampling process, the training data, especially the sampling weight to our BSDF importance sampling network, is guaranteed to be correct (unbiased), albeit some noise can remain. Therefore, the source of bias in our results can only originate from the learning process of the network itself, as will be analyzed in the next subsection.




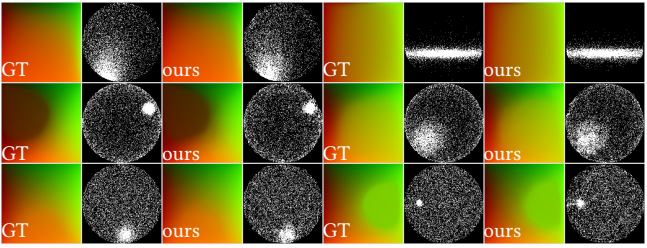
\begin{figure}[t]
    \centering
    \begin{overpic}[width=\columnwidth]{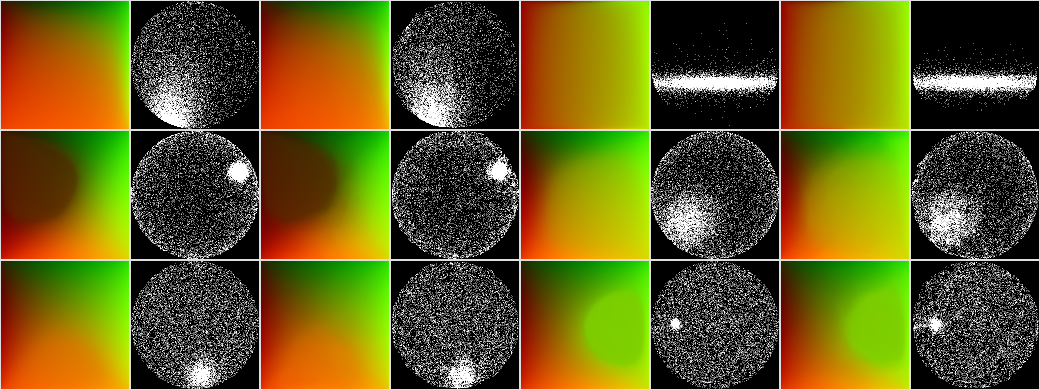}
    \put(0,1){\color{white}\small{GT}}
    \put(25,1){\color{white}\small{ours}}
    \put(50,1){\color{white}\small{GT}}
    \put(75,1){\color{white}\small{ours}}
    \put(0,13){\color{white}\small{GT}}
    \put(25,13){\color{white}\small{ours}}
    \put(50,13){\color{white}\small{GT}}
    \put(75,13){\color{white}\small{ours}}
    \put(0,26){\color{white}\small{GT}}
    \put(25,26){\color{white}\small{ours}}
    \put(50,26){\color{white}\small{GT}}
    \put(75,26){\color{white}\small{ours}}
    \put(0,39){\color{white}\small{GT}}
    \put(25,39){\color{white}\small{ours}}
    \put(50,39){\color{white}\small{GT}}
    \put(75,39){\color{white}\small{ours}}
    \end{overpic}   
\vspace{-6mm}
    \caption{We visualize pairs of importance maps and their corresponding binning results (overexposed to better visualize the difference) for different materials, including multiple-bounce microfacet conductor materials (top row), layered materials (middle tow) and Disney principled materials (bottom row), learned by our BSDF sampling network. By comparing with the ground truth, our model can accurately learn the importance maps and reproduce the BSDF lobes for various parametric BSDFs.    }
    \label{fig:sim_eval_sample}
\end{figure}


To validate the functionality of our BSDF sampling network, we compare the importance maps and the binning results between the ground truth and our network's prediction, as shown in Fig.~\ref{fig:sim_eval_sample}. Thanks to our use of optimal transport that results in smooth importance maps, our network is able to learn to compress them well.



So far, our solution to general parametric BSDF importance sampling has already completed. However, modern rendering pipeline usually supports multiple importance sampling (MIS). In the MIS framework, a renderer should implement not only BSDF sampling, but also BSDF evaluation (returning the BSDF value given the incident and outgoing directions) and PDF query (returning the PDF value similarly). For some BSDFs, these two tasks can be elegantly computed in closed form, e.g., the Disney Principled BSDFs. But for some other BSDFs, it is still difficult to evaluate the BSDF values, e.g., the layered BSDFs.

Therefore, \emph{for completeness}, we also compress the BSDFs and PDFs, using separate neural networks.


\paragraph{BSDF evaluation network}

For BSDF evaluation, our evaluation network $\mathcal{E}$ takes an additional incident direction $\bo_{i}$ and outputs the BSDF value as a 3-channel RGB value:
\begin{equation}
   \mathcal{E}(\epsilon, \bo_{o}, \bo_{i}) = f_s(\bo_{o},\bo_{i}) \langle\bo_{i},\bn\rangle.
\label{equ:eval}
\end{equation}

Note again that the BSDF evaluation network is \emph{optional} and should only be used when there is no analytical BSDF evaluation scheme. We choose the simple network structure because it fulfills our need already, producing correct BSDF values, as validated in Fig.~\ref{teaser} and Fig.~\ref{fig:comp_multiple_MIS}. We are aware of other existing neural BSDF/BTF compression methods \cite{Rainer2019Neural, Rainer2020Unified, sztrajman2021neural, Fan:2022:NLBRDF, zheng2021compact}. And we also believe that their success can further strengthen our importance sampling scheme when combined together, replacing our simple evaluation network. But we do not extend analysis on them any further, since BSDF evaluation is not our main contribution.

\paragraph{PDF query network} Our PDF query network $\mathcal{P}$ has a very similar definition compared to the BSDF evaluation network. It also takes in the combinations of BSDF parameters as well as the incident directions and returns the PDF value of sampling that direction under solid angle measurement:
\begin{equation}
   \mathcal{P}(\epsilon, \bo_{o}, \bo_{i}) = \mathrm{PDF}\left(f_s(\bo_{o},\bo_{i}) \langle\bo_{i},\bn\rangle\right), 
\label{equ:pdf}
\end{equation}

\begin{figure}
    \includegraphics[width = \columnwidth]{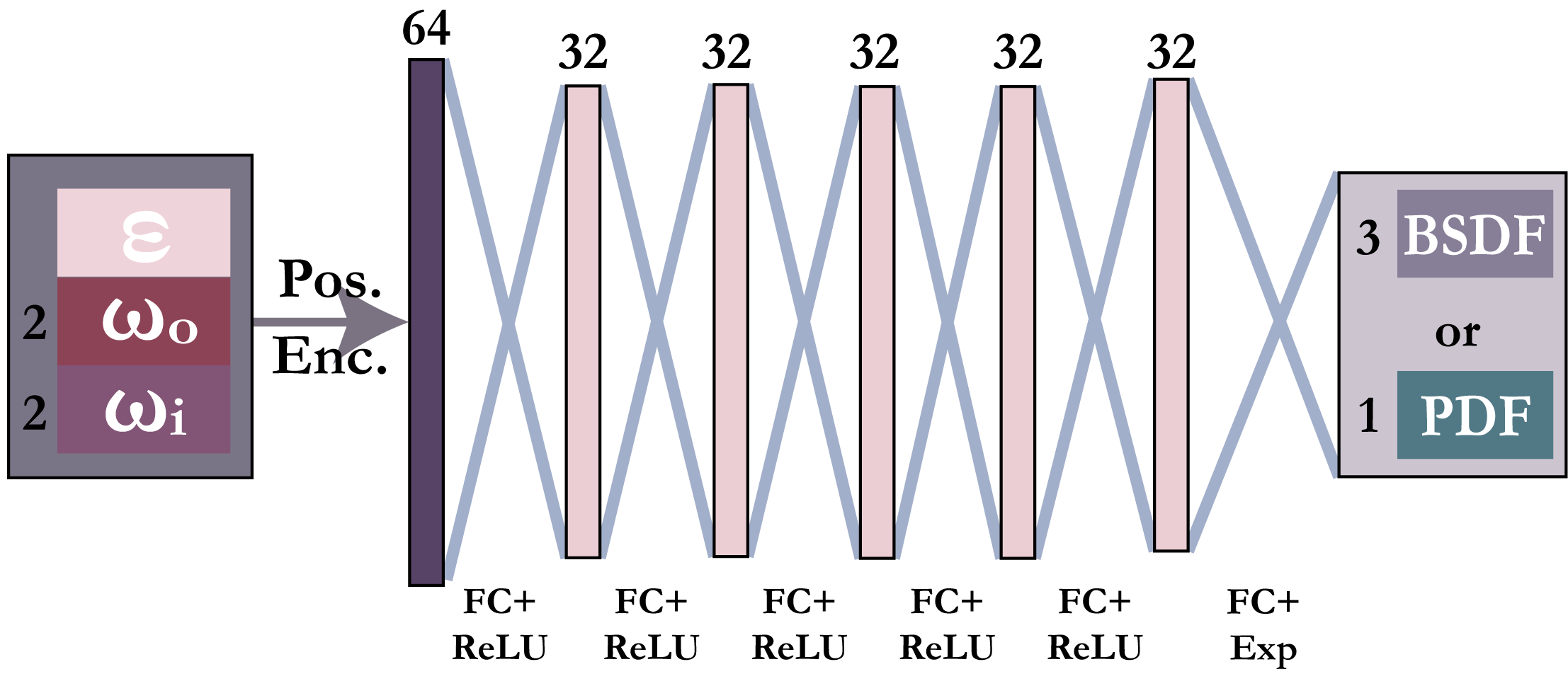}
    \caption{The structure of our BSDF evaluation and PDF query networks. These two networks take material parameters $\epsilon$, outgoing direction $\omega_{o}$, and incident direction $\omega_{i}$ as inputs and output the BSDF value or PDF, respectively. More details are shown in Sec.~\ref{sec:implemt_training}.}
    \label{fig:nn_eval_pdf}
\end{figure}

For readers unfamiliar with rendering, we make a special note here that our PDF query networks only serve as providing the PDF values for the computation of MIS weights (not sampling weights). A physically correct PDF is certainly welcomed -- as it will reduce the variance of the MIS combined results. However, even completely wrong PDFs in this step only leads to higher variance rather than any bias. We elaborate on this more in the next section with a simple but convincing experiment.



Our BSDF evaluation and PDF query networks share a similar lightweight structure, as illustrated in  Fig.~\ref{fig:nn_eval_pdf}. Note specifically that the emphasis on lightweight neural network design in core rendering~\cite{fur2022aggre,Fan:2022:NLBRDF} is different from that in deep learning. We use neural networks only as a general tool for efficiently compressing and querying high-dimensional data.

\subsection{Properties and analysis}
\label{sec:properties_analysis}

\paragraph{Source of Bias}
Bias is not desired, but there is no guarantee that neural networks produce fully unbiased results. In our case, the BSDF sampling network learns the sampling weight values, which is equivalent to  $\frac{f_{s}\langle\bo_{i},\bn\rangle}{\text{PDF}}$. However, this PDF is the correct PDF (rather than the output of our PDF query network) as part of the unbiased training data. And it will not be exposed to any other stages but only together as the sampling weight. Therefore, the only source of bias originates from the learning process of the network itself, and is on sampling weight only, which is usually low frequency and easy to learn as shown in Fig.~\ref{fig:sample_weight}.


\begin{figure}
    \begin{overpic}[width=\columnwidth]{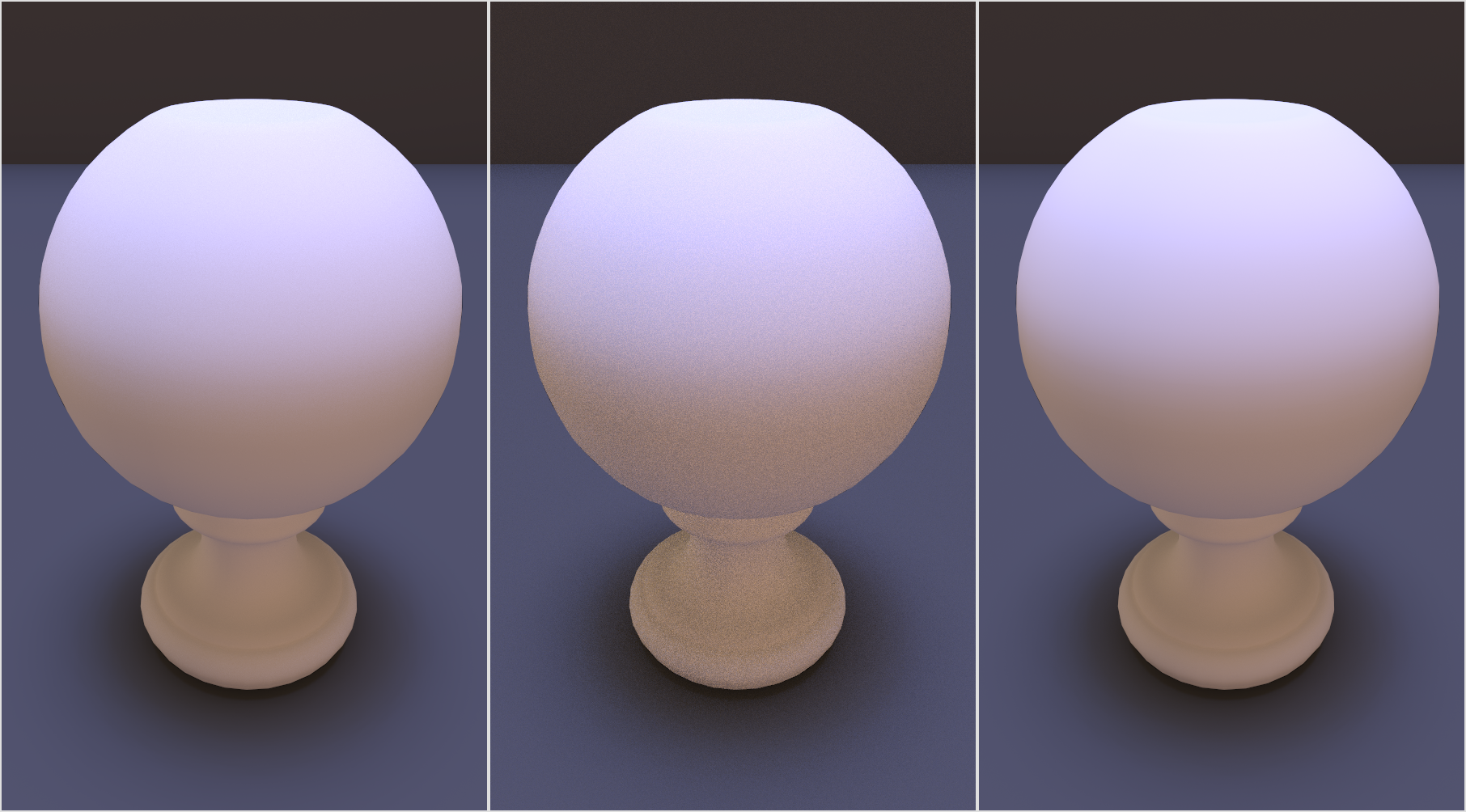}
    \put(1,1){\color{white}\small{$\mathbf{A}$}}
    \put(34,1){\color{white}\small{$\mathbf{B}$}}
    \put(67,1){\color{white}\small{$\mathbf{C}$}}
    \end{overpic}
    \vspace{-6mm}
    \caption{Rendered results of a ball status scene with a diffuse BSDF using different MIS PDFs: (a) -10, (b) tenth power of cos-weighted hemisphere PDF value, and (c) cos-weighted hemisphere PDF treated as the ground truth. All results are rendered with 16,384 samples for convergence. These three results are identical, which confirms that the MIS PDF value does not require any sum to one constraint, and the choice of MIS PDF does not introduce any bias.}
    \vspace{-4mm}
    \label{fig:MIS_PDF_comp}
\end{figure}

For our PDF query network, we do not include any sum-to-one (and other normalization) constraints. As mentioned earlier, this is because the PDF used in MIS can be any value, and the resulting MIS weights may even be negative \cite{kondapaneni2019optimal}. To further prove this statement, we show the MIS rendering results using a ball status scene with negative uniform MIS PDF, tenth power PDF, and cos-weighted PDF as reference in Fig.~\ref{fig:MIS_PDF_comp}.


\paragraph{Relationship to other BSDF sampling methods}

Our neural BSDF importance sampling solution has significant differences from other neural BSDF methods. For example, Fan et al.~\shortcite{Fan:2022:NLBRDF} sample different layered BRDF parameters by learning a Gaussian lobe and a Lambertian lobe. Similarly, Sztrajman et al.~\shortcite{sztrajman2021neural} apply a shallow network to learn the mapping between autoencoded BRDFs and parameters of fitted analytical Blinn-Phong importance sampling parameters. Though Sztrajman et al.~\shortcite{sztrajman2021neural} also use a neural network to calculate BRDF importance sampling, their network only provides fitting parameters to the analytical Blinn-Phong model.

\begin{figure}
    \begin{overpic}[width = \columnwidth]{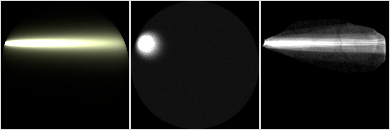}
    \put(1,1){\color{white}\small{BSDF slice}}
    \put(34,1){\color{white}\small{Fitting}}
    \put(67,1){\color{white}\small{Neural}}
    \end{overpic}
    \caption{ Binning result comparison between the fitted analytical Blinn-Phong model (middle) and our neural solution (right) on an anisotropic multiple-bounce microfacet BSDF slice. Our solution shows a better match than the fitted Blinn-Phong model to the reference BSDF slice.  }
    \label{fig:neural_fit}
\end{figure}

Therefore, the core of these importance sampling methods is still the analytical model fitting of BSDF slices. The insight of these strategies relies on a rough prior knowledge of the shape of the BSDF slices. When the prior breaks, the efficiency of sampling will be abysmal. For example, the Blinn-Phong importance sampling model serves as a good sample function for BSDFs with isotropic single specular lobe. However, anisotropic materials or complex multiple-lobe materials cannot be accurately described with only one specular highlight. We show the Blinn-Phong importance sampling fitting results of anisotropic multiple-bounce microfacet BSDF in Fig.~\ref{fig:neural_fit} for validation. 

We have differentiated our approach against normalizing flow-based importance sampling methods at the beginning of Sec.~\ref{sec:NetworkForImportanceBaking}, and more comparisons can be found in Fig.~\ref{fig:comp_multi_sample}. For other sampling methods, such as Dupuy et al.~\shortcite{dupuy2018adaptive}, they perform targeted design for specific kind of BSDFs. We consider them too far from our general parametric BSDF sampling, and do not extend the further discussion.


\section{Implementation details}

\subsection{Data Generation}
\label{sec:suppl_datagen}

\paragraph{BSDF slice generation}

We use an AMD 64-core 3995WX CPU machine for training data generation. The material parameters $\epsilon$ are sampled uniformly in their own space except the roughness is sampled in squared space. 
For outgoing directions, we sample them uniformly in square root of the polar angle to emphasize the grazing angle cases.
Each pair of $(\epsilon, \bo_{ox}, \bo_{oy})$ will be used for generating a 2D \emph{BSDF slice} with a resolution of 128$\times$128, where each pixel represents an incoming direction $(\bo_{ix}, \bo_{iy})$ and stores the three-channel $f_{s}\langle\bo_{i},\bn\rangle$ value. In our experiments, we generate 32,768 slices for our training on each material. 

The dataset for all the networks are generated from these 2D BSDF slices. These 2D BSDF slices can be directly used as training data for BSDF evaluation networks. For the PDF query network's dataset, we calculate the luminance of three-channel BSDFs to get single-channel data and then get the PDF by normalizing the luminance with the summed solid angle measurement. Finally, for the BSDF sampling network, we reparameterize the PDF from the hemisphere to a unit disk and then perform marginalized inverse transform sampling on the unit disk. This ensures that we are sampling on a correct distribution domain. The importance sampled sample points function as the target distributions of OT. Additionally, we also store the three-channel $\frac{f_{s}\langle\bo_{i},\bn\rangle}{\text{PDF}}$ values as the sampling weight. Note that the sampling and PDF calculation are all based on solid angle-measured BSDF radiance values, and there is no parameter space transformation. Therefore, there is no need for Jacobian calculation. In summary, all training data originate from BSDF evaluation.


\paragraph{Optimal transport}
In the next step, we use Geomloss \cite{feydy2019interpolating} to provide an initial mapping state, and then we use SOT \cite{paulin2020sliced} for further optimization. We choose to use two OT solutions because combining two methods will combine the advantages and avoid the limitations of these two methods. SOT computes swiftly, but we notice apparent crevices on \emph{BSDF slices} even after tens of thousands of iterations, especially for small roughness situations. The networks can grasp these crevices and produce incorrect dark areas in the rendering results. Geomloss runs slower than SOT, but it provides an outstanding initial state for SOT to optimize further. Note again, Geomloss and SOT are just ways of giving the mapping for the next learning step of our method. Other specific implementations of OT are all available choices.


\begin{table}[t]
\centering
    \begin{tabular}{c|cc}
\toprule[1pt]
Material & Parameter(Symbol) & Encoding Method \\ 
\hline
\multirowcell{5}{Multiple\\ Bounce\\ Material} & Roughness ($\alpha\in \mathbb{R}^2$) & $\text{freq}(\alpha) \in \mathbb{R}^{2\times 4}$ \\
                                 & $\text{R}_0$     ($a \in \mathbb{R}^3$) & $\text{freq}(a) \in \mathbb{R}^{3\times 4}$ \\
                                 & Camera Dir   ($\omega_{o} \in \mathbb{R}^2$) & $\text{freq}(\omega_{o}) \in \mathbb{R} ^ {2\times 12}$\\
                                 & Light Dir    ($\omega_{i} \in \mathbb{R}^2$) & $\text{freq}(\omega_{i}) \in \mathbb{R} ^ {2\times 12}$ \\ 
                                 & Sample Input      ($\xi \in \mathbb{R}^2$) & $\text{freq}(\xi) \in \mathbb{R} ^ {2\times 12}$ \\ 
\hline
\multirowcell{7}{Layered\\ Material}& Refractive Index ($\eta \in \mathbb{R}$)        &  $\text{ob}(\eta) \in \mathbb{R}^4$ \\
                                 &  Roughness           ($\alpha \in \mathbb{R}$)  &  $\text{ob}\left(\text{nl}(1 - e^{-\alpha})\right) \in \mathbb{R}^{4}$ \\
                                 & Attenuation Coeff ($\sigma_T \in \mathbb{R}$)   & $\text{ob}(\sigma_T) \in \mathbb{R}^5$ \\
                                 & Albedo                  ($a \in \mathbb{R}^3$) & $\text{id}(a) \in \mathbb{R}^{3}$ \\
                                 & Camera Dir     ($\omega_{o} \in \mathbb{R}^2$) & $\text{ob}(\omega_{o}) \in \mathbb{R} ^ {2\times 12}$\\
                                 & Light Dir      ($\omega_{i} \in \mathbb{R}^2$) & $\text{ob}(\omega_{i}) \in \mathbb{R} ^ {2\times 12}$ \\ 
                                 & Sample Input      ($\xi \in \mathbb{R}^2$) & $\text{ob}(\xi) \in \mathbb{R} ^ {2\times 12}$ \\ 
\hline
\multirowcell{5}{Disney\\ Material} & Metallic ($m \in \mathbb{R}$)        & $\text{ob}(m) \in \mathbb{R}^8$ \\
                                 &  Roughness           ($\alpha \in \mathbb{R}$)  &  $\text{ob}\left(\text{nl}(1 - e^{-\alpha})\right) \in \mathbb{R}^{8}$ \\
                                 & Camera Dir     ($\omega_{o} \in \mathbb{R}^2$) & $\text{ob}(\omega_{o}) \in \mathbb{R} ^ {2\times 12}$\\
                                 & Light Dir      ($\omega_{i} \in \mathbb{R}^2$) & $\text{ob}(\omega_{i}) \in \mathbb{R} ^ {2\times 12}$ \\ 
                                 & Sample Input      ($\xi \in \mathbb{R}^2$) & $\text{ob}(\xi) \in \mathbb{R} ^ {2\times 12}$ \\ 
\bottomrule[ 1pt]
\end{tabular}
\caption{ Parameters Encoding. For encoding method, freq($\cdot$) means frequency encoding \cite{mildenhall2020nerf} with both sine and cosine functions, ob($\cdot$) means one-blob encoding \cite{mueller2019neural}, id($\cdot$) stands for identity function, and nl($\cdot$) means normalize the value's range to $[0,1]$. }
\vspace{-6mm}
\label{table:encoding}
\end{table}

\subsection{Network and Training Details}
\label{sec:implemt_training}


Our networks are trained on a server with an Intel 16-core i9-7960X CPU and a NVIDIA 3090, and then the weights are saved into binary files for each kind of parametric BSDFs for renderer integration. 



\paragraph{Input Encoding}
Input encoding has been proved to provide sharper and more accurate results in several works \cite{mildenhall2020nerf, muller2021real}. Therefore, we adopt the input encoding for our query parameters similar to \cite{muller2021real}. The frequency encoding consists of $2k$ sine and cosine functions, with $k$ different frequencies from $2^0$ to $2^{k-1}$, and the one blob encoding \cite{mueller2019neural} has $l$ bins for the discrete Gaussian values of the encoded numbers. Details of each encoding for different types of materials are shown in Table~\ref{table:encoding}. 

\paragraph{Loss functions}

For BSDF evaluation networks, in order to capture both highlight area's intensity and non-highlight area's color, we use symmetric mean absolute percentage error (SMAPE) on BSDF values to optimize the networks as following:

$$
\mathcal{L}_{\text{eval}} = \dfrac{ \left\| \text{pred} - \text{gt} \right\|_1 }{\text{sg}(\left\| \text{pred} \right\|_1) + \text{sg}(\left\| \text{gt} \right\|_1) + \varepsilon}
$$

where $\varepsilon$ is set to $0.01$ and $\text{sg}(\cdot)$ means there is no gradient back propagation here. 

For PDF query networks, the predicted values are transformed into $\text{log}$ space and then we apply $\ell_1$ loss as follows:
$$
\mathcal{L}_{\text{pdf}} = \left\| \log (1+\text{pred}) - \log (1+\text{gt}) \right\|_1.
$$

For BSDF sampling network, we directly apply $\ell_1$ loss on network output, i.e. the incident direction. Additionally, to reduce the number of queries for networks in rendering, our network also returns the sampling weight. It is a low-frequency function (its visualization is shown in Fig.~\ref{fig:sample_weight}) that is averaged from the corresponding three-channel BSDFs. 
We use $\ell_1$ loss for the sampling weight outputs in $\log$ space. Thus, the full loss of our sample network shows below:
$$
\mathcal{L}_{\text{sample}} = \left\|  \hat{\omega} - \omega_{\text{gt}} \right\|_1 + \lambda\left\| \log (1+\hat{\text{sw}}) - \log (1+\text{sw}_{\text{gt}}) \right\|_1,
$$
where $\hat{\omega}$ indicates the predicted incident direction, and the $\hat{\text{sw}}$ is predicted sampling weight which is a 3 channel value. $\omega_{\text{gt}}$ and $\text{sw}_{\text{gt}}$ are the corresponding ground truth of them. We set $\lambda$ to $0.4$ in our training.

\paragraph{Training details}

We use the ADAM~\cite{kingma2014adam} optimizer with default parameters and cosine annealing scheduler to optimize our BSDF sampling networks, and Ranger~\cite{Ranger} optimizer to optimize the BSDF evaluation and PDF query networks. 
The batch size of training is set to $1048576$ and learning rate is set to $0.0001$. Each networks are trained with $500$ epochs in total. It takes about 48 hours to train BSDF sampling networks on each material, and 12 hours to train BSDF evaluation networks and PDF query networks separately on each material.


\subsection{Renderer Integration}
\label{sec:suppl_rendering}

We integrate our network in Mitsuba renderer \cite{jakob2010mitsuba} with minimum revisions to only BSDF classes. The network inferences are integrated with C++ and Mitsuba using Eigen \cite{eigenweb}. The fully-connected layers are interpreted as matrix multiplications implemented in three different classes. Inside neural BSDF classes, the evaluation, sample, and PDF functions will only need to call the network inference functions with corresponding parameters. At render time, the ray directions $\bo_{i}$ and $\bo_{o}$ for each shading point, as well as BSDF parameters $\epsilon$ serve as inputs of both BSDF evaluation and PDF query networks. The BSDF sampling network requires BSDF parameters $\epsilon$, outgoing direction $\bo_{o}$ and a pair of random numbers $(\xi_{0}, \xi_{1})$. Then the sampling network returns an incident direction $\bo_{i}$ together with the sampling weight. In this case, we do not need to refer to the BSDF evaluation and PDF query networks inside the sample function again. Using sampling weight relieves us from performing three network inferences and thus optimize the calculation efficiency. At the same time, it further decreases the bias brought about by the differences between BSDF sampling and PDF query networks. In sum, we have not made any revisions to the integrator. All the modifications we have made are only inside a neural BSDF class.
  
 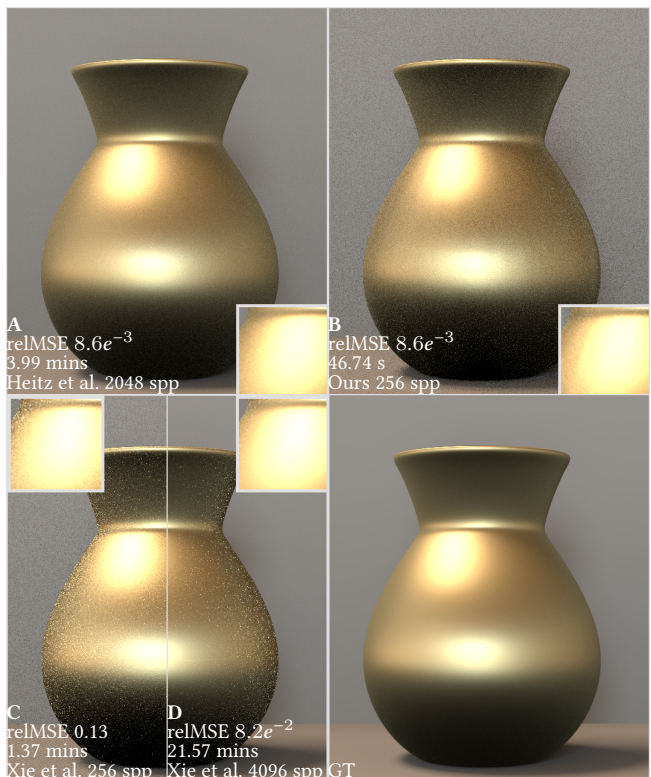
\begin{figure}[t]
    \centering
    \begin{subfigure}{\columnwidth}
    \centering
    \begin{overpic}[width=\columnwidth]{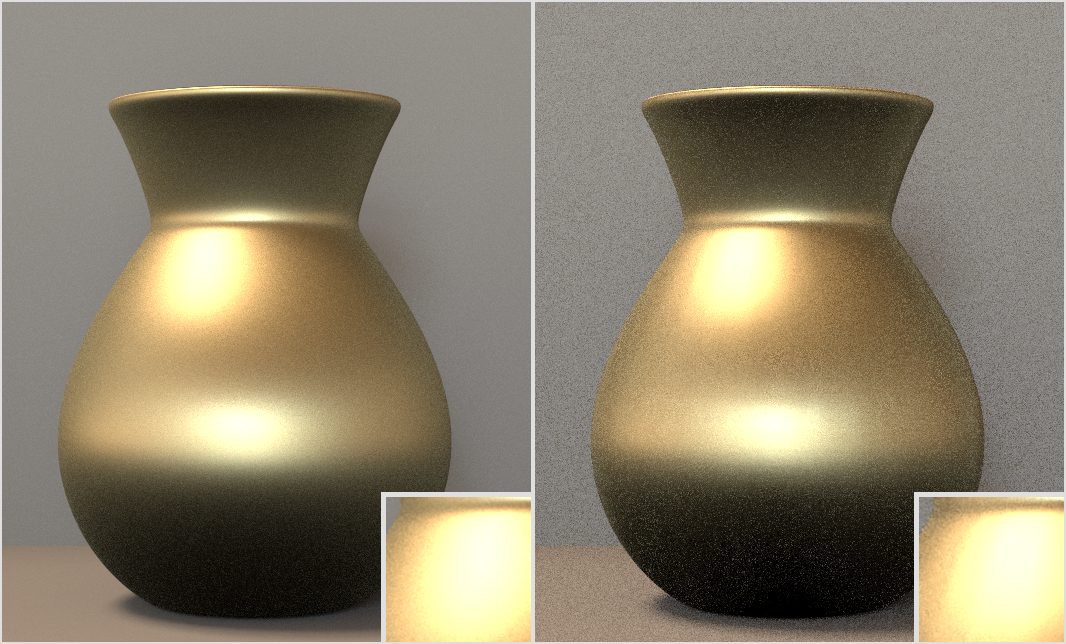}
    \put(0,1){\color{white}\small{Heitz et al. 2048 spp}}
    \put(0,4){\color{white}\small{3.99 mins}}
     \put(0,7){\color{white}\small{relMSE $8.6e^{-3}$}}
     \put(0,10){\color{white}\small{\textbf{A}}}
    \put(50,1){\color{white}\small{Ours 256 spp}}
    \put(50,4){\color{white}\small{46.74 s}}
     \put(50,7){\color{white}\small{relMSE $8.6e^{-3}$}}
     \put(50,10){\color{white}\small{\textbf{B}}}
    \end{overpic}
    \end{subfigure}
    \begin{subfigure}{\columnwidth}
    \centering
     \begin{overpic}[width=\columnwidth]{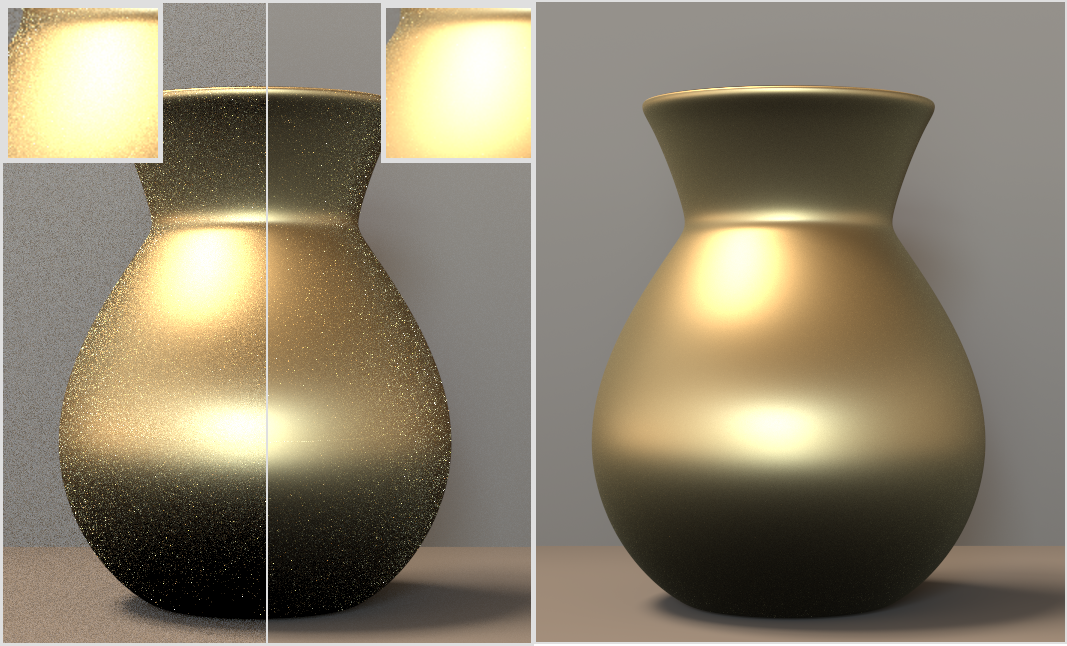}
    \put(0,1){\color{white}\small{Xie et al. 256 spp}}
    \put(0,4){\color{white}\small{1.37 mins}}
     \put(0,7){\color{white}\small{relMSE 0.13}}
     \put(0,10){\color{white}\small{\textbf{C}}}
    \put(25,1){\color{white}\small{Xie et al. 4096 spp}}
    \put(25,4){\color{white}\small{21.57 mins}}
     \put(25,7){\color{white}\small{relMSE $8.2e^{-2}$}}
     \put(25,10){\color{white}\small{\textbf{D}}}
    \put(50,1){\color{white}\small{GT}}
    \end{overpic}      
    \end{subfigure}
        \vspace{-4mm}
    \caption{We perform equal quality comparisons between (A) Heitz et al.~\shortcite{heitz2016multiple} and (B) our method (BSDF sampling network) on an anisotropic multiple-bounce microfacet conductor BSDF. Also, we perform equal-spp comparisons between (B) our method and (C) Xie et al. ~\shortcite{xie2019multiple}. Our method achieves less noise with better performance, while (D) Xie et al.~\shortcite{xie2019multiple} produce a noisy result with 16 times samples. Our method achieves the correct appearance and with less amount of time. }
    \label{fig:comp_multi_sample}
        \vspace{-4mm}
\end{figure}

\section{Results and Comparison}
\label{sec:results}

  We integrate our network in Mitsuba renderer \cite{jakob2010mitsuba}, and network inferences are integrated using C++ and Eigen \cite{eigenweb}. All the rendering performance is measured on an Intel 8-core i9-9900K machine. Now we only perform CPU network inference. 
 
 In this section, we validate our results on multiple-bounce rough conductors with the GGX NDF, position-free layered BSDFs, and Disney principled materials. Also, we use relative mean square error (relMSE) to measure the difference with the ground truth. Since our method and the other methods might converge to different ground truths, we use their own converged results as the ground truth (GT).

 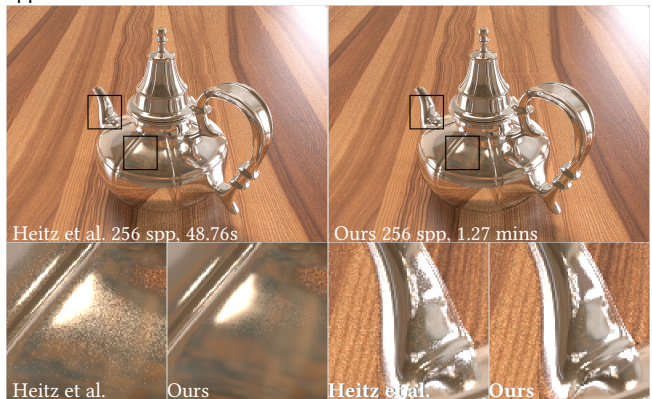
\begin{figure}[t]
\begin{subfigure}{\columnwidth}
    \begin{overpic}[width = \columnwidth]{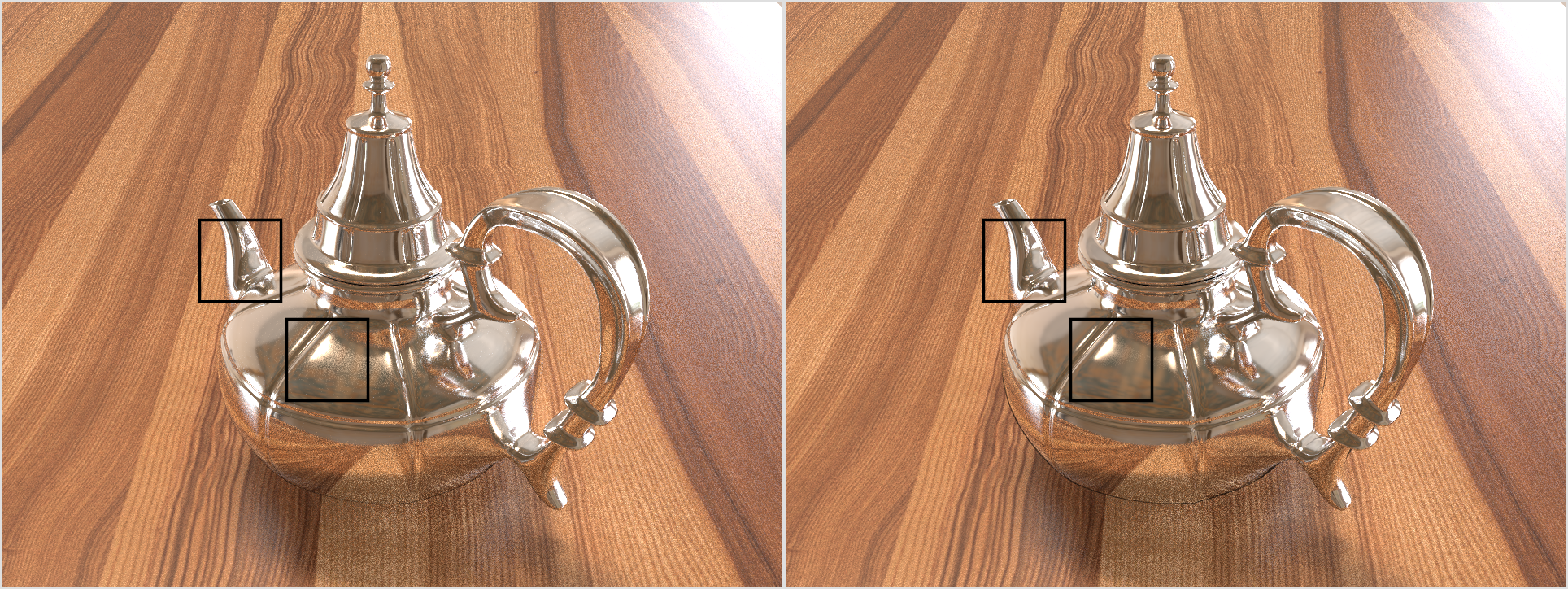}
    \put(1,1){\color{white}\small{Heitz et al. 256 spp, 48.76s}}
    \put(51,1){\color{white}\small{Ours 256 spp, 1.27 mins}}
    \end{overpic}
\end{subfigure}
\begin{subfigure}{\columnwidth}
    \begin{overpic}[width = \columnwidth]{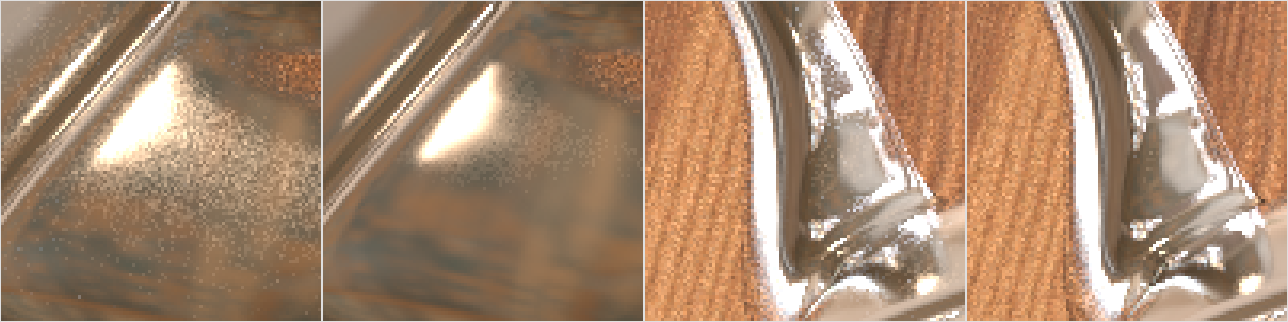}
    \put(1,1){\color{white}\small{Heitz et al.}}
    \put(25,1){\color{white}\small{Ours}}
    \put(50.2,1){\color{gray}\small{\textbf{Heitz et al.}}}
    \put(50,1){\color{white}\small{\textbf{Heitz et al.}}}
    \put(75,1){\color{white}\small{\textbf{Ours}}}
    \end{overpic}
\end{subfigure}
        \vspace{-4mm}
        \caption{Comparison between our method (BSDF sampling network) and Heitz et al. \shortcite{heitz2016multiple} on low-roughness ($\alpha = 0.01$) multiple-bounce microfacet conductor BSDFs with equal sampling rate. Our method shows correct highlights, while preserving a lower noise level than Heitz et al. \shortcite{heitz2016multiple}.}
    \label{fig:result_low_roughness}
\end{figure}

\subsection{Multiple-bounce Microfacet BRDFs}

For multiple-bounce microfacet conductors, we compare our rendering results with Heitz et al. \shortcite{heitz2016multiple} and use their method to generate all ground truth (GT) results. We start with comparing the efficiency and qualities of BSDF sampling only rendering results between our neural network and RealNVP used by Xie et al.~\shortcite{xie2019multiple}. We also perform comparisons for low-roughness conductors, as well as MIS rendering results. 

\paragraph{Vase scene} 
In Fig.~\ref{fig:comp_multi_sample}, we show an anisotropic conductor vase with $\alpha_{x} = 0.3$ and $\alpha_{y} = 0.1$ under an environment light, considering direct lighting only. We use our BSDF sampling network trained on the multiple-bounce dataset and compare it against Heitz et al.~\shortcite{heitz2016multiple} and Xie et al.~ \shortcite{xie2019multiple}. For fairness, we set the hidden dimension of the RealNVP network by Xie et al.~ \shortcite{xie2019multiple} as 32, keeping a similar parameter size as ours, and train their model at the same time as ours. Since their model only provides the sampled direction and its PDF, we call the GT evaluation function and divide it by their PDF as the sampling weights. For all the methods, we perform BSDF sampling only. Our method use less time to achieve equal quality than Heitz et al.~\shortcite{heitz2016multiple}. At the same time, the result by Xie et al.~ \shortcite{xie2019multiple} has higher variance with $2\times$ time cost.


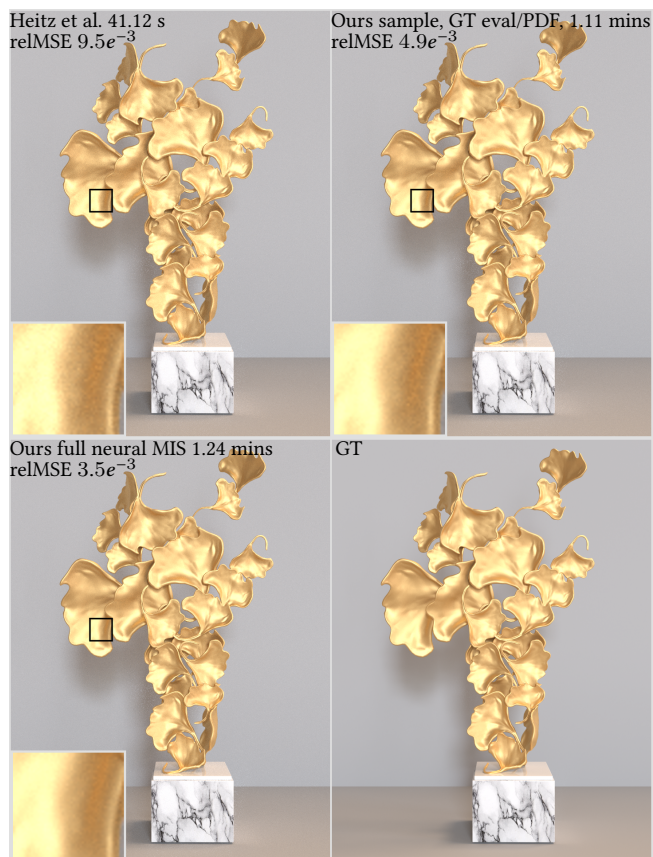
\begin{figure}[t]
\centering
    \begin{overpic}[width=\columnwidth]{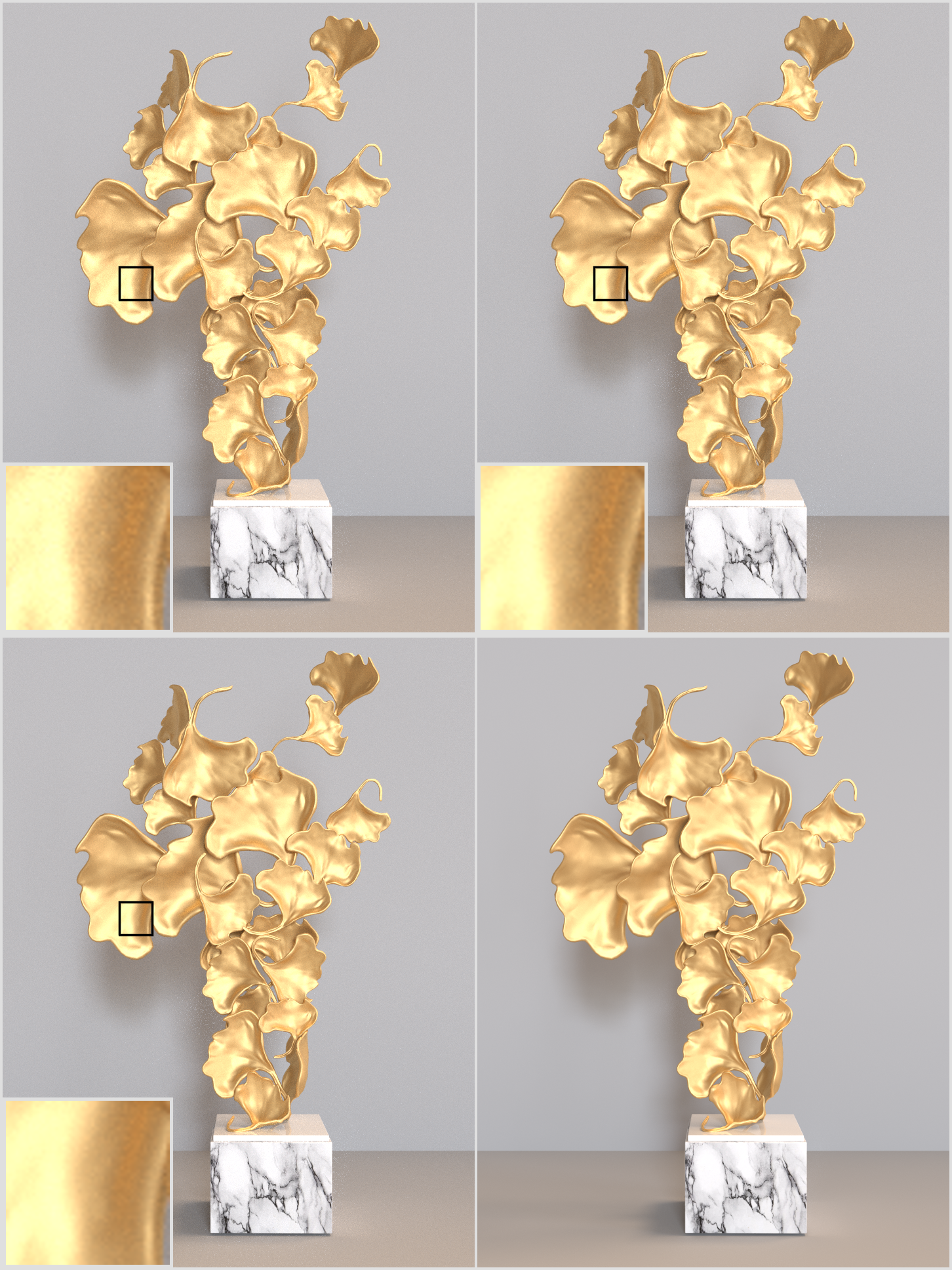}
    \put(0,48){\color{black}\small{Ours full neural MIS 1.24 mins}}
    \put(0,45.5){\color{black}\small{relMSE $3.5e^{-3}$}}
  \put(38,48){\color{black}\small{GT}}
    \put(0,98){\color{black}\small{Heitz et al. 41.12 s}}
    \put(0,95.5){\color{black}\small{relMSE $9.5e^{-3}$}}
  \put(37.5,98){\color{black}\small{Ours sample, GT eval/PDF, 1.11 mins}}
     \put(37.5,95.5){\color{black}\small{relMSE $4.9e^{-3}$}}
    \end{overpic}
        \vspace{-8mm}
    \caption{Comparison between our method (BSDF sampling network + GT evaluation and PDF), our method (three networks), Heitz et al.~\shortcite{heitz2016multiple} and the ground truth rendered with 256 spp. The roughness is set as 0.3. By introducing our BSDF sampling network only, our model is already able to produce results with lower variance, while the other components (evaluation and PDF query network) further reduce the noise level, at the cost of a longer time.}
        \vspace{-4mm}
    \label{fig:comp_multiple_MIS}
\end{figure}

\paragraph{Low roughness teapot scene.} 
BRDF sampling is also tricky for multiple-bounce microfacet conductors with low roughness, due to the high frequency of the BRDF lobes. To show the effectiveness of our BSDF sampling network on such cases, we compare our method against Heitz et al.~\shortcite{heitz2016multiple} on the \emph{teapot scene} (with $\alpha = 0.01$) with equal-spp in Fig.~\ref{fig:result_low_roughness}. In this comparison, we perform BSDF sampling only. Our result preserves the correct highlights and shows much less variance than Heitz et al.~\shortcite{heitz2016multiple}, with only $0.5 \times$ extra cost.

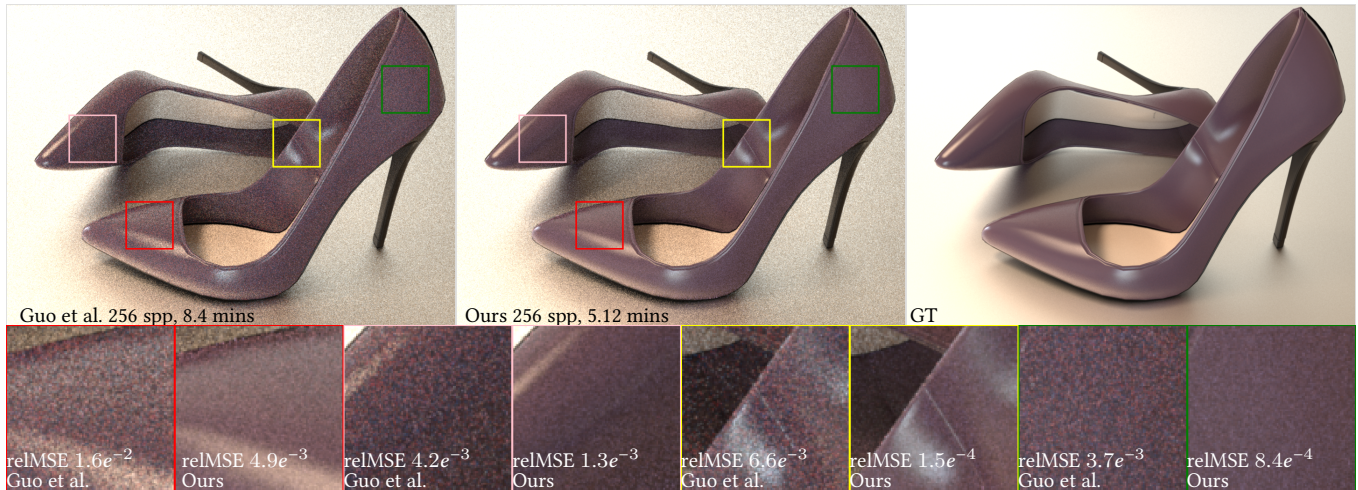
\begin{figure*}[t]
\begin{subfigure}{\textwidth}
    \begin{overpic}[width = \columnwidth]{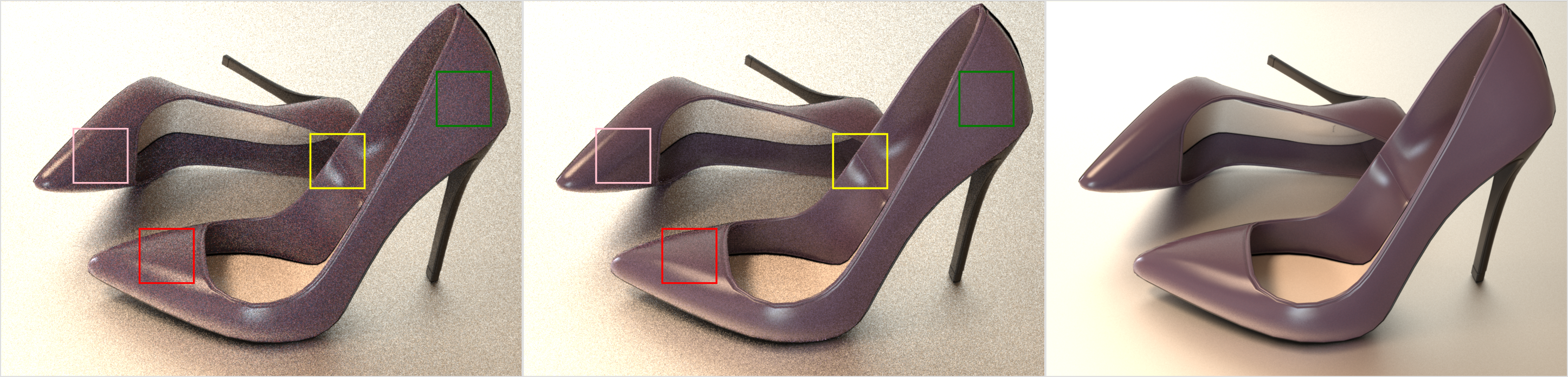}
    \put(1,0.5){\color{black}\small{Guo et al. 256 spp, 8.4 mins}}
    \put(34,0.5){\color{black}\small{Ours 256 spp, 5.12 mins}}
    \put(67,0.5){\color{black}\small{GT}}
    \end{overpic}
\end{subfigure}
\begin{subfigure}{\textwidth}
    \begin{overpic}[width = \columnwidth]{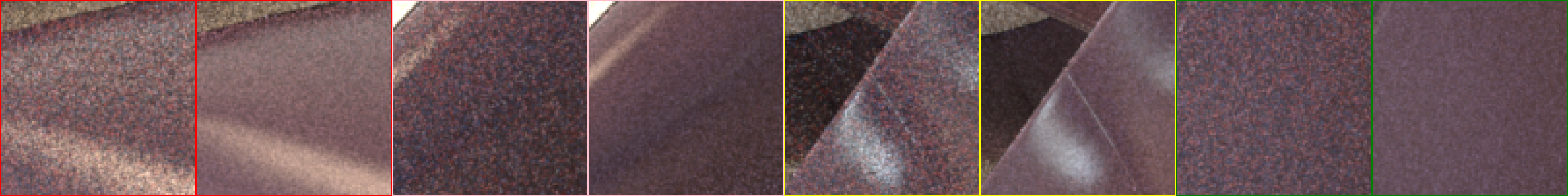}
    \put(0,0.5){\color{white}\small{Guo et al.}}
    \put(0,2){\color{white}\small{relMSE $1.6e^{-2}$}}
    \put(13,0.5){\color{white}\small{Ours}}
    \put(13,2){\color{white}\small{relMSE $4.9e^{-3}$}}    
    \put(25,0.5){\color{white}\small{Guo et al.}}
    \put(25,2){\color{white}\small{relMSE $4.2e^{-3}$}}   
    \put(37.5,0.5){\color{white}\small{Ours}}
    \put(37.5,2){\color{white}\small{relMSE $1.3e^{-3}$}}   
    \put(50,0.5){\color{white}\small{Guo et al.}}
    \put(50,2){\color{white}\small{relMSE $6.6e^{-3}$}}   
    \put(62.5,0.5){\color{white}\small{Ours}}
    \put(62.5,2){\color{white}\small{relMSE $1.5e^{-4}$}}   
    \put(75,0.5){\color{white}\small{Guo et al.}}
    \put(75,2){\color{white}\small{relMSE $3.7e^{-3}$}}   
    \put(87.5,0.5){\color{white}\small{Ours}}
    \put(87.5,2){\color{white}\small{relMSE $8.4e^{-4}$}}   
    \end{overpic}
\end{subfigure}
        \vspace{-4mm}
    \caption{Equal-spp comparison between our model (BSDF sampling network) and Guo et al. ~\shortcite{guo2018position} on a layered material. Our method not only shows less variance but also has higher efficiency. The layered material consists of a dense media in the middle, leading to an expensive random walk for Guo et al.~\shortcite{guo2018position}. }
        \vspace{-3mm}
    \label{fig:result_layered_shoe_SW}
\end{figure*}

\paragraph{Ginkgo ornament scene} Besides the BSDF sampling network, we also introduce BSDF evaluation and PDF query networks, which are essential for complex BSDFs with heavy evaluations, like the multiple-bounce Smith microfacet BSDFs. Our full solution enables MIS for Monte Carlo rendering. In Fig.~\ref{fig:comp_multiple_MIS}, we compare our methods (sample network only and full solution), and Heitz et al.~\shortcite{heitz2016multiple}. For our method (sample network only), we use the BSDF evaluation and PDF from Heitz et al.~\shortcite{heitz2016multiple} to enable MIS. By comparison, we find that our full solution outperforms the others in terms of rendering quality, with only a slight time overhead.

\paragraph{Conductor Kitchen Shelf} In Fig.~\ref{teaser} (top), we compare our full solution (three networks) with Heitz et al.~\shortcite{heitz2016multiple} on a variety of objects with different colors and roughness lit by an area light. The roughness of the four objects (from left to right) are set as $\alpha= 0.2$, $\alpha = 0.6$, $\alpha = 0.03$ and $\alpha_{x} = 0.08$ and $\alpha_{y} = 0.3$, covering from low roughness to high roughness, from isotropic to anisotropic materials. By comparison,  our result shows higher quality than Heitz et al.~\shortcite{heitz2016multiple} both visually and quantitatively. However, the time cost of our method is about 2$\times$ slower, due to the network inference. Since our current network inference is a simple CPU implementation, we believe that further optimization can decrease our time cost significantly.

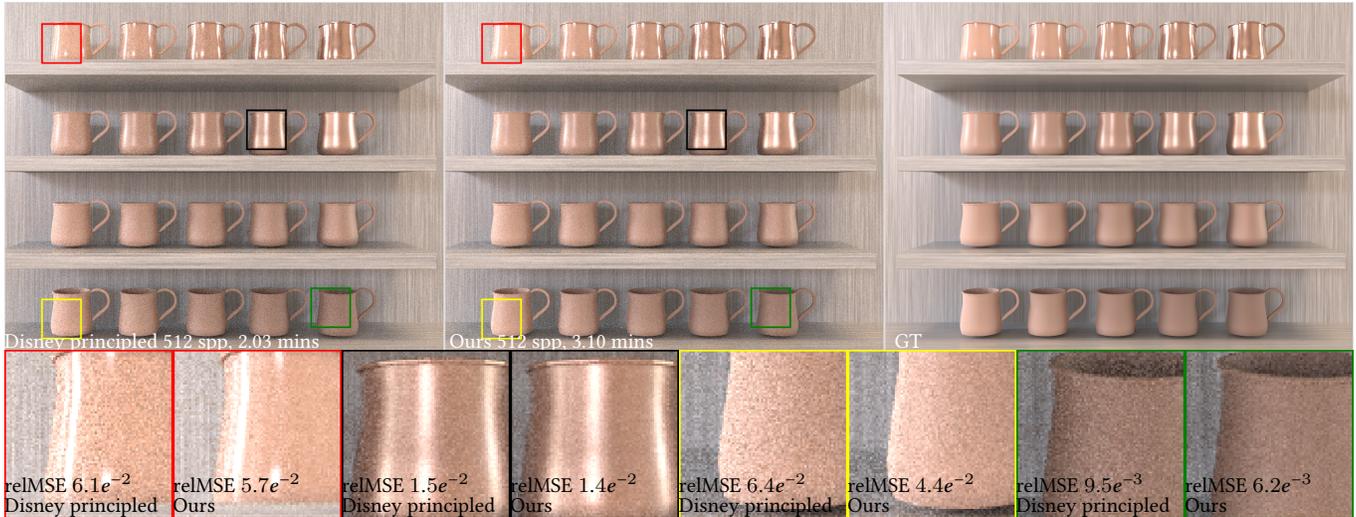
\begin{figure*}[!]
\begin{subfigure}{\textwidth}
    \begin{overpic}[width = \columnwidth]{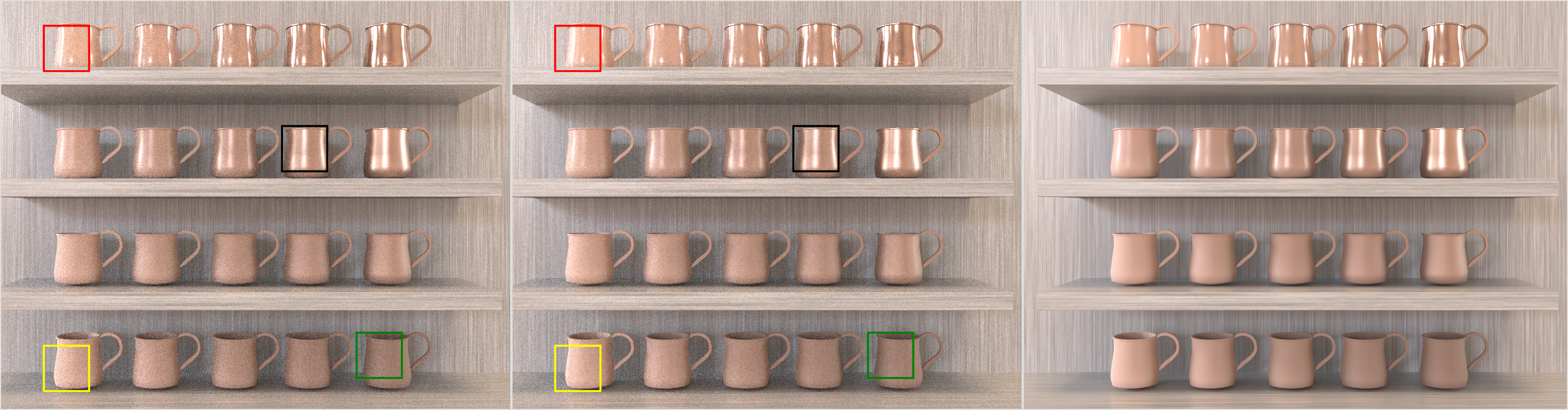}
    \put(0,0.5){\color{white}\small{Disney principled 512 spp, 2.03 mins}}
    \put(33,0.5){\color{white}\small{Ours 512 spp, 3.10 mins}}
    \put(66,0.5){\color{white}\small{GT}}
      \end{overpic}
\end{subfigure}
\begin{subfigure}{\textwidth}
     \begin{overpic}[width = \columnwidth]{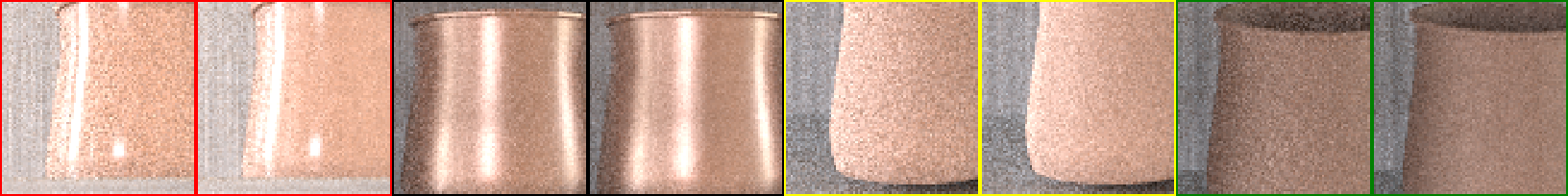}
    \put(0,0.5){\color{black}\small{Disney principled}}
    \put(0,2){\color{black}\small{relMSE $6.1e^{-2}$}}   
    \put(12.5,0.5){\color{black}\small{Ours}}
    \put(12.5,2){\color{black}\small{relMSE $5.7e^{-2}$}}   
    \put(25,0.5){\color{black}\small{Disney principled}}
    \put(25,2){\color{black}\small{relMSE $1.5e^{-2}$}}   
    \put(37.5,0.5){\color{black}\small{Ours}}
    \put(37.5,2){\color{black}\small{relMSE $1.4e^{-2}$}}   
    \put(50,0.5){\color{black}\small{Disney principled}}
    \put(50,2){\color{black}\small{relMSE $6.4e^{-2}$}}   
    \put(62.5,0.5){\color{black}\small{Ours}}
    \put(62.5,2){\color{black}\small{relMSE $4.4e^{-2}$}}   
    \put(75,0.5){\color{black}\small{Disney principled}}
    \put(75,2){\color{black}\small{relMSE $9.5e^{-3}$}}   
    \put(87.5,0.5){\color{black}\small{Ours}}
    \put(87.5,2){\color{black}\small{relMSE $6.2e^{-3}$}}   
  \end{overpic}
\end{subfigure}
        \vspace{-4mm}
        \caption{Equal-spp comparison between our model (BSDF sampling network) and Disney Principled BSDF on a broad range of materials, with metallic  $\mathbf{m}$, varies from 0.1, 0.3, 0.5, 0.7, to 0.9, and roughness $\alpha$ varies from 0.15, 0.35, 0.65, to 0.85. Our method shows a lower noise level than Disney Principled BSDFs, proving that our method provides a better BSDF importance sampling strategy than theirs. }
        \vspace{-3mm}
    \label{fig:result_disney}
\end{figure*}

\subsection{Position-free Layered BSDFs}

In this section, we demonstrate effectiveness of our model on layered materials by comparing with Guo et al.~\shortcite{guo2018position} and use this method to generate ground truth (GT). Again, we only show results on two-layer BSDFs for simplicity, and more layers can be handled in the same way. 

\paragraph{Layered Shoes} 

In Fig.~\ref{fig:result_layered_shoe_SW}, we compare our BSDF sampling model (trained on layered materials) with Guo et al.~\shortcite{guo2018position} with equal spp on a Shoe scene lit by an environment map. The shoes have a dielectric material with roughness $\alpha = 0.08$ and refractive index $\eta = 1.5$ as the top layer, a diffuse with color as the substrate, together with a medium ($\sigma_{T} = 0.8$). Both methods perform BSDF sampling only, without MIS. With an equal sampling rate, the result by Guo et al.~\shortcite{guo2018position} is much noisier while having a much longer time than ours. The expensive time cost of Guo et al.~\shortcite{guo2018position} is due to the random walk in the dense medium. 

\paragraph{Layered Kitchen Shelf}
Then, we validate our full solution (three networks) on layered BSDFs, by comparing against Guo et al.~\shortcite{guo2018position} with equal time in Fig.~\ref{teaser} (bottom). In this scene, we show several objects, with different roughness $\alpha$, refractive index $\eta$, attenuation coefficient $\sigma_{T}$, and albedo color lit by an environment map, considering both direct lighting and indirect lighting. Both methods use MIS to render this scene. To achieve equal time, we use 1024 spp for our method and 850 spp for Guo et al.~\shortcite{guo2018position}. Our result has a lower variance and higher computational efficiency than theirs.

\subsection{Disney Principled BSDFs}
Next, we show some results on Disney Principled BSDFs. We choose to vary two parameters \emph{metallic} and \emph{roughness} while fixing the remaining parameters. As a comparison, we implement full Disney principled model in Mitsuba renderer to generate ground truth (GT) results. In Fig.~\ref{fig:result_disney}, we show an equal-spp comparison between our model (BSDF sampling model) and Disney principled BSDF on the coffee cup scene with varying parameters (metallic and roughness), lit by an environment map. Both our model and Disney principled BSDF use BSDF sampling only. Under all these settings, our method achieves less noise level than Disney Principled BSDFs, especially for materials with low metallic and roughness, at the cost of a longer rendering time (1.5$\times$) due to the network inference. It proves our method provides a better BSDF importance sampling strategy than Disney Principled BSDFs.

\section{Discussion and Limitations}
\label{sec:discussion}

\paragraph{Unified representation of BSDF evaluation and sampling} In our BSDF evaluation network, when the BSDF parameters are specified, one outgoing direction $\bo_{o}$ will produce a corresponding 2D BSDF slice, where each pixel is a BSDF value. This BSDF slice is queried using the 2D incident direction $\bo_{i}$. Similarly, given the BSDF parameters and an outgoing direction, a 2D importance map is established, where a pixel represents the mapped sample position and the sampling weight. This importance map is queried with a 2D random number $(\xi_0,\xi_1)$. 
With this similarity identified, we have the evidence to claim that BSDF evaluation and sampling (and PDF query) are in essence very similar. Therefore, training a neural network for sampling should not be more complex than training for evaluation. This observation also shows that prohibitively expensive neural network structures, such as RealNVP and NICE, can be avoided.



\begin{figure}[b]
    \begin{overpic}[width = \columnwidth]{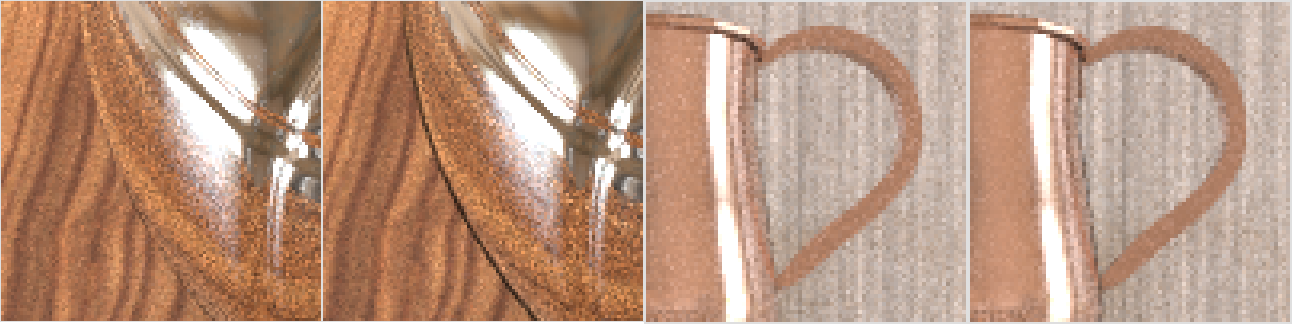}
    \put(0,0.5){\color{black}\small{Heitz et al.}}
    \put(25,0.5){\color{black}\small{Ours}}
    \put(50,0.5){\color{black}\small{Disney principled}}
    \put(75,0.5){\color{black}\small{Ours}}
      \end{overpic}   
      \caption{Failure case. Our method introduces bias at grazing angles for highly specular materials. For example, the edges of the teapot and the coffee cup have an apparent black border. }
    \label{fig:limitation_fail}
\end{figure}
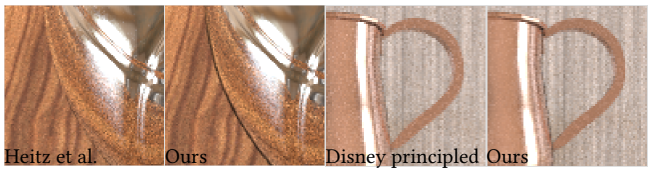

\paragraph{Non-parametric/measured BSDFs.} 
Intuitively, non-parametric BSDFs are often large data blocks. However, they are in fact of much lower dimensions than the parametric BSDFs. For example, a measured bidirectional texture function (BTF) has a much lower-dimensional parameter space (6D) than any of our examples (see Table.~\ref{table:encoding}). Besides, in our work, we focus on BSDF importance sampling rather than BSDF evaluation, and we have already analyzed commonly used sampling schemes for measured BSDFs, such as fitting and normalizing flow.


\paragraph{Performance.} Our results correctly reconstruct the parametric BSDFs' appearance with comparatively less noise, but the performance influence cannot be neglected even though our networks are small. Since we simply use inline CPU integration to fully integrate our neural networks into the renderer with minimum revision to the rendering pipeline, the inference of our neural networks is far from optimized. There are many ways to further improve inference efficiency, such as using a GPU inference framework, e.g., TensorRT, and/or devoting to considerable engineering optimization~\cite{muller2021real}.


\paragraph{Bias.} Neural network prediction will inevitably introduce bias even though the original training data are unbiased. Strict applications such as white furnace test will expose the bias issue immediately. Our method does have visible bias, as show in  Fig.~\ref{fig:limitation_fail}, but we do not see apparent problems in these results from our practical applications. Nevertheless, we believe it is still meaningful to study other methods that enable unbiased importance map compression. 


\section{Conclusion and Future Work}

We have introduced BSDF importance baking, a lightweight neural solution to perform perfect importance sampling of parametric BSDFs. We start from the observation that the mapping that performs importance sampling on a BSDF slice can be simply recorded as a 2D importance map. Following this observation, we propose to use optimal transport to precompute the importance maps accurately; then, we use a lightweight neural network to compress them efficiently. Together with an optional BSDF evaluation network and PDF query network, our method enables full multiple importance sampling (MIS) without any revision to the rendering pipeline. Compared with previous methods, we demonstrate reduced noise levels on rendering results with a rich set of appearances, including conductors with anisotropic roughness, layered BSDF, and Disney principled materials.

We believe that we have brought about novel contributions: our method is the first to utilize optimal transport in rendering applications that are not affected by its heavy computation during runtime; and our method is the first complete neural alternative with the potential to fully replace parametric BSDFs with MIS.

In the future, an immediate research direction is to improve the quality of our lightweight neural networks and further improve their performance. Sampling other forms of appearance representation could also be interesting, for example, using our importance baking scheme to investigate the sampling problem of 4D light fields data or 5D neural radiance fields (NeRF) data. Apart from neural compression, other data compression strategies could be explored as well, until a more efficient or unbiased method is found.



%
%
%
%

\bibliographystyle{ACM-Reference-Format}
\bibliography{paper}


\begin{thebibliography}{48}


\ifx \showCODEN    \undefined \def \showCODEN     #1{\unskip}     \fi
\ifx \showDOI      \undefined \def \showDOI       #1{#1}\fi
\ifx \showISBNx    \undefined \def \showISBNx     #1{\unskip}     \fi
\ifx \showISBNxiii \undefined \def \showISBNxiii  #1{\unskip}     \fi
\ifx \showISSN     \undefined \def \showISSN      #1{\unskip}     \fi
\ifx \showLCCN     \undefined \def \showLCCN      #1{\unskip}     \fi
\ifx \shownote     \undefined \def \shownote      #1{#1}          \fi
\ifx \showarticletitle \undefined \def \showarticletitle #1{#1}   \fi
\ifx \showURL      \undefined \def \showURL       {\relax}        \fi
\providecommand\bibfield[2]{#2}
\providecommand\bibinfo[2]{#2}
\providecommand\natexlab[1]{#1}
\providecommand\showeprint[2][]{arXiv:#2}

\bibitem[Ashikhmin and Shirley(2001)]%
        {Ashikhmin2001}
\bibfield{author}{\bibinfo{person}{Michael Ashikhmin} {and}
  \bibinfo{person}{Peter Shirley}.} \bibinfo{year}{2001}\natexlab{}.
\newblock \showarticletitle{An Anisotropic Phong Light Reflection Model}.
\newblock \bibinfo{journal}{\emph{Journal of Graphics Tools}}
  \bibinfo{volume}{5} (\bibinfo{date}{01} \bibinfo{year}{2001}).
\newblock


\bibitem[Beckmann and Spizzichino(1963)]%
        {BeckmannSpizzichino:1963}
\bibfield{author}{\bibinfo{person}{P. Beckmann} {and} \bibinfo{person}{A.
  Spizzichino}.} \bibinfo{year}{1963}\natexlab{}.
\newblock \bibinfo{booktitle}{\emph{The scattering of electromagnetic waves
  from rough surfaces}}.
\newblock \bibinfo{publisher}{Pergamon Press.}
\newblock


\bibitem[Blinn(1977)]%
        {blinn1977models}
\bibfield{author}{\bibinfo{person}{James~F Blinn}.}
  \bibinfo{year}{1977}\natexlab{}.
\newblock \showarticletitle{Models of light reflection for computer synthesized
  pictures}. In \bibinfo{booktitle}{\emph{Proceedings of the 4th annual
  conference on Computer graphics and interactive techniques}}.
  \bibinfo{pages}{192--198}.
\newblock


\bibitem[Bonneel and Coeurjolly(2019)]%
        {BC19}
\bibfield{author}{\bibinfo{person}{Nicolas Bonneel} {and}
  \bibinfo{person}{David Coeurjolly}.} \bibinfo{year}{2019}\natexlab{}.
\newblock \showarticletitle{SPOT: Sliced Partial Optimal Transport}.
\newblock \bibinfo{journal}{\emph{ACM Transactions on Graphics (SIGGRAPH)}}
  \bibinfo{volume}{38}, \bibinfo{number}{4} (\bibinfo{year}{2019}).
\newblock


\bibitem[Bonneel et~al\mbox{.}(2016)]%
        {bonneel2016wasserstein}
\bibfield{author}{\bibinfo{person}{Nicolas Bonneel}, \bibinfo{person}{Gabriel
  Peyr{\'e}}, {and} \bibinfo{person}{Marco Cuturi}.}
  \bibinfo{year}{2016}\natexlab{}.
\newblock \showarticletitle{Wasserstein barycentric coordinates: histogram
  regression using optimal transport.}
\newblock \bibinfo{journal}{\emph{ACM Trans. Graph.}} \bibinfo{volume}{35},
  \bibinfo{number}{4} (\bibinfo{year}{2016}), \bibinfo{pages}{71--1}.
\newblock


\bibitem[Bonneel et~al\mbox{.}(2011)]%
        {bonneel2011displacement}
\bibfield{author}{\bibinfo{person}{Nicolas Bonneel}, \bibinfo{person}{Michiel
  Van De~Panne}, \bibinfo{person}{Sylvain Paris}, {and}
  \bibinfo{person}{Wolfgang Heidrich}.} \bibinfo{year}{2011}\natexlab{}.
\newblock \showarticletitle{Displacement interpolation using Lagrangian mass
  transport}. In \bibinfo{booktitle}{\emph{Proceedings of the 2011 SIGGRAPH
  Asia conference}}. \bibinfo{pages}{1--12}.
\newblock


\bibitem[Burley(2012)]%
        {burley2012physically}
\bibfield{author}{\bibinfo{person}{Brent Burley}.}
  \bibinfo{year}{2012}\natexlab{}.
\newblock \showarticletitle{Physically-based shading at disney}. In
  \bibinfo{booktitle}{\emph{ACM SIGGRAPH}}, Vol.~\bibinfo{volume}{2012}. vol.
  2012, \bibinfo{pages}{1--7}.
\newblock


\bibitem[Clarberg et~al\mbox{.}(2005)]%
        {clarberg2005wavelet}
\bibfield{author}{\bibinfo{person}{Petrik Clarberg}, \bibinfo{person}{Wojciech
  Jarosz}, \bibinfo{person}{Tomas Akenine-M{\"o}ller}, {and}
  \bibinfo{person}{Henrik~Wann Jensen}.} \bibinfo{year}{2005}\natexlab{}.
\newblock \showarticletitle{Wavelet importance sampling: efficiently evaluating
  products of complex functions}.
\newblock In \bibinfo{booktitle}{\emph{ACM SIGGRAPH 2005 Papers}}.
  \bibinfo{pages}{1166--1175}.
\newblock


\bibitem[Cuturi(2013)]%
        {cuturi2013sinkhorn}
\bibfield{author}{\bibinfo{person}{Marco Cuturi}.}
  \bibinfo{year}{2013}\natexlab{}.
\newblock \showarticletitle{Sinkhorn distances: Lightspeed computation of
  optimal transport}.
\newblock \bibinfo{journal}{\emph{Advances in neural information processing
  systems}}  \bibinfo{volume}{26} (\bibinfo{year}{2013}).
\newblock


\bibitem[Dinh et~al\mbox{.}(2014)]%
        {dinh2014nice}
\bibfield{author}{\bibinfo{person}{Laurent Dinh}, \bibinfo{person}{David
  Krueger}, {and} \bibinfo{person}{Yoshua Bengio}.}
  \bibinfo{year}{2014}\natexlab{}.
\newblock \showarticletitle{Nice: Non-linear independent components
  estimation}.
\newblock \bibinfo{journal}{\emph{arXiv preprint arXiv:1410.8516}}
  (\bibinfo{year}{2014}).
\newblock


\bibitem[Dinh et~al\mbox{.}(2016)]%
        {dinh2016density}
\bibfield{author}{\bibinfo{person}{Laurent Dinh}, \bibinfo{person}{Jascha
  Sohl-Dickstein}, {and} \bibinfo{person}{Samy Bengio}.}
  \bibinfo{year}{2016}\natexlab{}.
\newblock \showarticletitle{Density estimation using real nvp}.
\newblock \bibinfo{journal}{\emph{arXiv preprint arXiv:1605.08803}}
  (\bibinfo{year}{2016}).
\newblock


\bibitem[Dupuy and Jakob(2018)]%
        {dupuy2018adaptive}
\bibfield{author}{\bibinfo{person}{Jonathan Dupuy} {and}
  \bibinfo{person}{Wenzel Jakob}.} \bibinfo{year}{2018}\natexlab{}.
\newblock \showarticletitle{An adaptive parameterization for efficient material
  acquisition and rendering}.
\newblock \bibinfo{journal}{\emph{ACM Transactions on graphics (TOG)}}
  \bibinfo{volume}{37}, \bibinfo{number}{6} (\bibinfo{year}{2018}),
  \bibinfo{pages}{1--14}.
\newblock


\bibitem[Fan et~al\mbox{.}(2021)]%
        {fan2021neural}
\bibfield{author}{\bibinfo{person}{Jiahui Fan}, \bibinfo{person}{Beibei Wang},
  \bibinfo{person}{Milo{\v{s}} Ha{\v{s}}an}, \bibinfo{person}{Jian Yang}, {and}
  \bibinfo{person}{Ling-Qi Yan}.} \bibinfo{year}{2021}\natexlab{}.
\newblock \showarticletitle{Neural BRDFs: Representation and Operations}.
\newblock \bibinfo{journal}{\emph{arXiv preprint arXiv:2111.03797}}
  (\bibinfo{year}{2021}).
\newblock


\bibitem[Fan et~al\mbox{.}(2022)]%
        {Fan:2022:NLBRDF}
\bibfield{author}{\bibinfo{person}{Jiahui Fan}, \bibinfo{person}{Beibei Wang},
  \bibinfo{person}{Milo\v{s} Ha\v{s}an}, \bibinfo{person}{Jian Yang}, {and}
  \bibinfo{person}{Ling-Qi Yan}.} \bibinfo{year}{2022}\natexlab{}.
\newblock \showarticletitle{Neural Layered BRDFs}. In
  \bibinfo{booktitle}{\emph{Proceedings of SIGGRAPH 2022}}.
\newblock


\bibitem[Feydy et~al\mbox{.}(2019)]%
        {feydy2019interpolating}
\bibfield{author}{\bibinfo{person}{Jean Feydy}, \bibinfo{person}{Thibault
  S{\'e}journ{\'e}}, \bibinfo{person}{Fran{\c{c}}ois-Xavier Vialard},
  \bibinfo{person}{Shun-ichi Amari}, \bibinfo{person}{Alain Trouve}, {and}
  \bibinfo{person}{Gabriel Peyr{\'e}}.} \bibinfo{year}{2019}\natexlab{}.
\newblock \showarticletitle{Interpolating between Optimal Transport and MMD
  using Sinkhorn Divergences}. In \bibinfo{booktitle}{\emph{The 22nd
  International Conference on Artificial Intelligence and Statistics}}.
  \bibinfo{pages}{2681--2690}.
\newblock


\bibitem[Guennebaud et~al\mbox{.}(2010)]%
        {eigenweb}
\bibfield{author}{\bibinfo{person}{Ga\"{e}l Guennebaud},
  \bibinfo{person}{Beno\^{i}t Jacob}, {et~al\mbox{.}}}
  \bibinfo{year}{2010}\natexlab{}.
\newblock \bibinfo{title}{Eigen v3}.
\newblock \bibinfo{howpublished}{http://eigen.tuxfamily.org}.
\newblock


\bibitem[Guo et~al\mbox{.}(2018)]%
        {guo2018position}
\bibfield{author}{\bibinfo{person}{Yu Guo}, \bibinfo{person}{Milo{\v{s}}
  Ha{\v{s}}an}, {and} \bibinfo{person}{Shuang Zhao}.}
  \bibinfo{year}{2018}\natexlab{}.
\newblock \showarticletitle{Position-free Monte Carlo simulation for arbitrary
  layered BSDFs}.
\newblock \bibinfo{journal}{\emph{ACM Transactions on Graphics (ToG)}}
  \bibinfo{volume}{37}, \bibinfo{number}{6} (\bibinfo{year}{2018}),
  \bibinfo{pages}{1--14}.
\newblock


\bibitem[Heitz(2017)]%
        {heitz2017simpler}
\bibfield{author}{\bibinfo{person}{Eric Heitz}.}
  \bibinfo{year}{2017}\natexlab{}.
\newblock \emph{\bibinfo{title}{A Simpler and Exact Sampling Routine for the
  GGX Distribution of Visible Normals}}.
\newblock \bibinfo{thesistype}{Ph.\,D. Dissertation}. \bibinfo{school}{Unity
  Technologies}.
\newblock


\bibitem[Heitz(2018)]%
        {heitz2018sampling}
\bibfield{author}{\bibinfo{person}{Eric Heitz}.}
  \bibinfo{year}{2018}\natexlab{}.
\newblock \showarticletitle{Sampling the GGX distribution of visible normals}.
\newblock \bibinfo{journal}{\emph{Journal of Computer Graphics Techniques}}
  \bibinfo{volume}{7}, \bibinfo{number}{4} (\bibinfo{year}{2018}),
  \bibinfo{pages}{1--13}.
\newblock


\bibitem[Heitz and d'Eon(2014)]%
        {heitz2014importance}
\bibfield{author}{\bibinfo{person}{Eric Heitz} {and} \bibinfo{person}{Eugene
  d'Eon}.} \bibinfo{year}{2014}\natexlab{}.
\newblock \showarticletitle{Importance sampling microfacet-based BSDFs using
  the distribution of visible normals}. In \bibinfo{booktitle}{\emph{Computer
  Graphics Forum}}, Vol.~\bibinfo{volume}{33}. Wiley Online Library,
  \bibinfo{pages}{103--112}.
\newblock


\bibitem[Heitz et~al\mbox{.}(2016)]%
        {heitz2016multiple}
\bibfield{author}{\bibinfo{person}{Eric Heitz}, \bibinfo{person}{Johannes
  Hanika}, \bibinfo{person}{Eugene d'Eon}, {and} \bibinfo{person}{Carsten
  Dachsbacher}.} \bibinfo{year}{2016}\natexlab{}.
\newblock \showarticletitle{Multiple-scattering microfacet BSDFs with the Smith
  model}.
\newblock \bibinfo{journal}{\emph{ACM Transactions on Graphics (TOG)}}
  \bibinfo{volume}{35}, \bibinfo{number}{4} (\bibinfo{year}{2016}),
  \bibinfo{pages}{1--14}.
\newblock


\bibitem[Hu et~al\mbox{.}(2020)]%
        {hu2020deepbrdf}
\bibfield{author}{\bibinfo{person}{Bingyang Hu}, \bibinfo{person}{Jie Guo},
  \bibinfo{person}{Yanjun Chen}, \bibinfo{person}{Mengtian Li}, {and}
  \bibinfo{person}{Yanwen Guo}.} \bibinfo{year}{2020}\natexlab{}.
\newblock \showarticletitle{DeepBRDF: A Deep Representation for Manipulating
  Measured BRDF}. In \bibinfo{booktitle}{\emph{Computer Graphics Forum}},
  Vol.~\bibinfo{volume}{39}. Wiley Online Library, \bibinfo{pages}{157--166}.
\newblock


\bibitem[Jakob(2010)]%
        {jakob2010mitsuba}
\bibfield{author}{\bibinfo{person}{Wenzel Jakob}.}
  \bibinfo{year}{2010}\natexlab{}.
\newblock \bibinfo{title}{Mitsuba renderer}.
\newblock
\newblock


\bibitem[Kantorovich(1942)]%
        {kantorovich1942translocation}
\bibfield{author}{\bibinfo{person}{Leonid~V Kantorovich}.}
  \bibinfo{year}{1942}\natexlab{}.
\newblock \showarticletitle{On the translocation of masses}. In
  \bibinfo{booktitle}{\emph{Dokl. Akad. Nauk. USSR (NS)}},
  Vol.~\bibinfo{volume}{37}. \bibinfo{pages}{199--201}.
\newblock


\bibitem[Kim et~al\mbox{.}(2008)]%
        {kim2008feature}
\bibfield{author}{\bibinfo{person}{Dongyeon Kim}, \bibinfo{person}{Minjung
  Son}, \bibinfo{person}{Yunjin Lee}, \bibinfo{person}{Henry Kang}, {and}
  \bibinfo{person}{Seungyong Lee}.} \bibinfo{year}{2008}\natexlab{}.
\newblock \showarticletitle{Feature-guided image stippling}. In
  \bibinfo{booktitle}{\emph{Computer Graphics Forum}},
  Vol.~\bibinfo{volume}{27}. Wiley Online Library, \bibinfo{pages}{1209--1216}.
\newblock


\bibitem[Kingma and Ba(2014)]%
        {kingma2014adam}
\bibfield{author}{\bibinfo{person}{Diederik~P Kingma} {and}
  \bibinfo{person}{Jimmy Ba}.} \bibinfo{year}{2014}\natexlab{}.
\newblock \showarticletitle{Adam: A method for stochastic optimization}.
\newblock \bibinfo{journal}{\emph{arXiv preprint arXiv:1412.6980}}
  (\bibinfo{year}{2014}).
\newblock


\bibitem[Kondapaneni et~al\mbox{.}(2019)]%
        {kondapaneni2019optimal}
\bibfield{author}{\bibinfo{person}{Ivo Kondapaneni}, \bibinfo{person}{Petr
  V{\'e}voda}, \bibinfo{person}{Pascal Grittmann},
  \bibinfo{person}{Tom{\'a}{\v{s}} Sk{\v{r}}ivan}, \bibinfo{person}{Philipp
  Slusallek}, {and} \bibinfo{person}{Jaroslav K{\v{r}}iv{\'a}nek}.}
  \bibinfo{year}{2019}\natexlab{}.
\newblock \showarticletitle{Optimal multiple importance sampling}.
\newblock \bibinfo{journal}{\emph{ACM Transactions on Graphics (TOG)}}
  \bibinfo{volume}{38}, \bibinfo{number}{4} (\bibinfo{year}{2019}),
  \bibinfo{pages}{1--14}.
\newblock


\bibitem[Kuznetsov et~al\mbox{.}(2021)]%
        {kuznetsov2021neumip}
\bibfield{author}{\bibinfo{person}{Alexandr Kuznetsov},
  \bibinfo{person}{Krishna Mullia}, \bibinfo{person}{Zexiang Xu},
  \bibinfo{person}{Milo\v{s} Ha\v{s}an}, {and} \bibinfo{person}{Ravi
  Ramamoorthi}.} \bibinfo{year}{2021}\natexlab{}.
\newblock \showarticletitle{NeuMIP: Multi-Resolution Neural Materials}.
\newblock \bibinfo{journal}{\emph{Transactions on Graphics (Proceedings of
  SIGGRAPH)}} \bibinfo{volume}{40}, \bibinfo{number}{4}, Article
  \bibinfo{articleno}{175} (\bibinfo{date}{July} \bibinfo{year}{2021}),
  \bibinfo{numpages}{13}~pages.
\newblock


\bibitem[Lawrence et~al\mbox{.}(2004)]%
        {lawrence2004efficient}
\bibfield{author}{\bibinfo{person}{Jason Lawrence}, \bibinfo{person}{Szymon
  Rusinkiewicz}, {and} \bibinfo{person}{Ravi Ramamoorthi}.}
  \bibinfo{year}{2004}\natexlab{}.
\newblock \showarticletitle{Efficient BRDF importance sampling using a factored
  representation}.
\newblock \bibinfo{journal}{\emph{ACM Transactions on Graphics (ToG)}}
  \bibinfo{volume}{23}, \bibinfo{number}{3} (\bibinfo{year}{2004}),
  \bibinfo{pages}{496--505}.
\newblock


\bibitem[McAuley et~al\mbox{.}(2012)]%
        {mcauley2012practical}
\bibfield{author}{\bibinfo{person}{Stephen McAuley}, \bibinfo{person}{Stephen
  Hill}, \bibinfo{person}{Naty Hoffman}, \bibinfo{person}{Yoshiharu Gotanda},
  \bibinfo{person}{Brian Smits}, \bibinfo{person}{Brent Burley}, {and}
  \bibinfo{person}{Adam Martinez}.} \bibinfo{year}{2012}\natexlab{}.
\newblock \showarticletitle{Practical physically-based shading in film and game
  production}.
\newblock In \bibinfo{booktitle}{\emph{ACM SIGGRAPH 2012 Courses}}.
  \bibinfo{pages}{1--7}.
\newblock


\bibitem[Mildenhall et~al\mbox{.}(2020)]%
        {mildenhall2020nerf}
\bibfield{author}{\bibinfo{person}{Ben Mildenhall}, \bibinfo{person}{Pratul~P.
  Srinivasan}, \bibinfo{person}{Matthew Tancik}, \bibinfo{person}{Jonathan~T.
  Barron}, \bibinfo{person}{Ravi Ramamoorthi}, {and} \bibinfo{person}{Ren Ng}.}
  \bibinfo{year}{2020}\natexlab{}.
\newblock \showarticletitle{NeRF: Representing Scenes as Neural Radiance Fields
  for View Synthesis}. In \bibinfo{booktitle}{\emph{ECCV}}.
\newblock


\bibitem[Monge(1781)]%
        {monge1781memoire}
\bibfield{author}{\bibinfo{person}{Gaspard Monge}.}
  \bibinfo{year}{1781}\natexlab{}.
\newblock \showarticletitle{M{\'e}moire sur la th{\'e}orie des d{\'e}blais et
  des remblais}.
\newblock \bibinfo{journal}{\emph{Histoire de l'Acad{\'e}mie Royale des
  Sciences de Paris}} (\bibinfo{year}{1781}).
\newblock


\bibitem[M{\"u}ller et~al\mbox{.}(2019)]%
        {Muller18}
\bibfield{author}{\bibinfo{person}{Thomas M{\"u}ller}, \bibinfo{person}{Brian
  McWilliams}, \bibinfo{person}{Fabrice Rousselle}, \bibinfo{person}{Markus
  Gross}, {and} \bibinfo{person}{Jan Nov{\'a}k}.}
  \bibinfo{year}{2019}\natexlab{}.
\newblock \showarticletitle{Neural importance sampling}.
\newblock \bibinfo{journal}{\emph{ACM Transactions on Graphics}}
  \bibinfo{volume}{38}, \bibinfo{number}{5} (\bibinfo{year}{2019}).
\newblock


\bibitem[M\"{u}ller et~al\mbox{.}(2019)]%
        {mueller2019neural}
\bibfield{author}{\bibinfo{person}{Thomas M\"{u}ller}, \bibinfo{person}{Brian
  McWilliams}, \bibinfo{person}{Fabrice Rousselle}, \bibinfo{person}{Markus
  Gross}, {and} \bibinfo{person}{Jan Nov\'{a}k}.}
  \bibinfo{year}{2019}\natexlab{}.
\newblock \showarticletitle{Neural Importance Sampling}.
\newblock \bibinfo{journal}{\emph{ACM Trans. Graph.}} \bibinfo{volume}{38},
  \bibinfo{number}{5}, Article \bibinfo{articleno}{145} (\bibinfo{date}{Oct.}
  \bibinfo{year}{2019}), \bibinfo{numpages}{19}~pages.
\newblock
\showISSN{0730-0301}
\urldef\tempurl%
\url{https://doi.org/10.1145/3341156}
\showDOI{\tempurl}


\bibitem[M{\"u}ller et~al\mbox{.}(2021)]%
        {muller2021real}
\bibfield{author}{\bibinfo{person}{Thomas M{\"u}ller}, \bibinfo{person}{Fabrice
  Rousselle}, \bibinfo{person}{Jan Nov{\'a}k}, {and} \bibinfo{person}{Alexander
  Keller}.} \bibinfo{year}{2021}\natexlab{}.
\newblock \showarticletitle{Real-time neural radiance caching for path
  tracing}.
\newblock \bibinfo{journal}{\emph{arXiv preprint arXiv:2106.12372}}
  (\bibinfo{year}{2021}).
\newblock


\bibitem[Paulin et~al\mbox{.}(2020)]%
        {paulin2020sliced}
\bibfield{author}{\bibinfo{person}{Lois Paulin}, \bibinfo{person}{Nicolas
  Bonneel}, \bibinfo{person}{David Coeurjolly}, \bibinfo{person}{Jean-Claude
  Iehl}, \bibinfo{person}{Antoine Webanck}, \bibinfo{person}{Mathieu Desbrun},
  {and} \bibinfo{person}{Victor Ostromoukhov}.}
  \bibinfo{year}{2020}\natexlab{}.
\newblock \showarticletitle{Sliced optimal transport sampling.}
\newblock \bibinfo{journal}{\emph{ACM Trans. Graph.}} \bibinfo{volume}{39},
  \bibinfo{number}{4} (\bibinfo{year}{2020}), \bibinfo{pages}{99}.
\newblock


\bibitem[Phong(1975)]%
        {phong1975}
\bibfield{author}{\bibinfo{person}{Bui~Tuong Phong}.}
  \bibinfo{year}{1975}\natexlab{}.
\newblock \showarticletitle{Illumination for Computer Generated Pictures}.
\newblock \bibinfo{journal}{\emph{Commun. ACM}} \bibinfo{volume}{18},
  \bibinfo{number}{6} (\bibinfo{date}{jun} \bibinfo{year}{1975}),
  \bibinfo{pages}{311–317}.
\newblock
\showISSN{0001-0782}
\urldef\tempurl%
\url{https://doi.org/10.1145/360825.360839}
\showDOI{\tempurl}


\bibitem[Rainer et~al\mbox{.}(2020)]%
        {Rainer2020Unified}
\bibfield{author}{\bibinfo{person}{Gilles Rainer}, \bibinfo{person}{Abhijeet
  Ghosh}, \bibinfo{person}{Wenzel Jakob}, {and} \bibinfo{person}{Tim Weyrich}.}
  \bibinfo{year}{2020}\natexlab{}.
\newblock \showarticletitle{Unified Neural Encoding of BTFs}.
\newblock \bibinfo{journal}{\emph{Computer Graphics Forum (Proceedings of
  Eurographics)}} \bibinfo{volume}{39}, \bibinfo{number}{2}
  (\bibinfo{date}{June} \bibinfo{year}{2020}).
\newblock
\urldef\tempurl%
\url{https://doi.org/10.1111/cgf.13921}
\showDOI{\tempurl}


\bibitem[Rainer et~al\mbox{.}(2019)]%
        {Rainer2019Neural}
\bibfield{author}{\bibinfo{person}{Gilles Rainer}, \bibinfo{person}{Wenzel
  Jakob}, \bibinfo{person}{Abhijeet Ghosh}, {and} \bibinfo{person}{Tim
  Weyrich}.} \bibinfo{year}{2019}\natexlab{}.
\newblock \showarticletitle{Neural BTF Compression and Interpolation}.
\newblock \bibinfo{journal}{\emph{Computer Graphics Forum (Proceedings of
  Eurographics)}} \bibinfo{volume}{38}, \bibinfo{number}{2}
  (\bibinfo{date}{March} \bibinfo{year}{2019}).
\newblock


\bibitem[Solomon et~al\mbox{.}(2015)]%
        {solomon2015convolutional}
\bibfield{author}{\bibinfo{person}{Justin Solomon}, \bibinfo{person}{Fernando
  De~Goes}, \bibinfo{person}{Gabriel Peyr{\'e}}, \bibinfo{person}{Marco
  Cuturi}, \bibinfo{person}{Adrian Butscher}, \bibinfo{person}{Andy Nguyen},
  \bibinfo{person}{Tao Du}, {and} \bibinfo{person}{Leonidas Guibas}.}
  \bibinfo{year}{2015}\natexlab{}.
\newblock \showarticletitle{Convolutional wasserstein distances: Efficient
  optimal transportation on geometric domains}.
\newblock \bibinfo{journal}{\emph{ACM Transactions on Graphics (ToG)}}
  \bibinfo{volume}{34}, \bibinfo{number}{4} (\bibinfo{year}{2015}),
  \bibinfo{pages}{1--11}.
\newblock


\bibitem[Sztrajman et~al\mbox{.}(2021)]%
        {sztrajman2021neural}
\bibfield{author}{\bibinfo{person}{Alejandro Sztrajman},
  \bibinfo{person}{Gilles Rainer}, \bibinfo{person}{Tobias Ritschel}, {and}
  \bibinfo{person}{Tim Weyrich}.} \bibinfo{year}{2021}\natexlab{}.
\newblock \showarticletitle{Neural BRDF Representation and Importance
  Sampling}. In \bibinfo{booktitle}{\emph{Computer Graphics Forum}},
  Vol.~\bibinfo{volume}{40}. Wiley Online Library, \bibinfo{pages}{332--346}.
\newblock


\bibitem[Walter et~al\mbox{.}(2007)]%
        {walter2007microfacet}
\bibfield{author}{\bibinfo{person}{Bruce Walter}, \bibinfo{person}{Stephen~R
  Marschner}, \bibinfo{person}{Hongsong Li}, {and} \bibinfo{person}{Kenneth~E
  Torrance}.} \bibinfo{year}{2007}\natexlab{}.
\newblock \showarticletitle{Microfacet Models for Refraction through Rough
  Surfaces.}
\newblock \bibinfo{journal}{\emph{Rendering techniques}}
  \bibinfo{volume}{2007} (\bibinfo{year}{2007}), \bibinfo{pages}{18th}.
\newblock


\bibitem[Wang et~al\mbox{.}(2022)]%
        {wang2022position}
\bibfield{author}{\bibinfo{person}{Beibei Wang}, \bibinfo{person}{Wenhua Jin},
  \bibinfo{person}{Jiahui Fan}, \bibinfo{person}{Jian Yang},
  \bibinfo{person}{Nicolas Holzschuch}, {and} \bibinfo{person}{Ling-Qi Yan}.}
  \bibinfo{year}{2022}\natexlab{}.
\newblock \showarticletitle{Position-free multiple-bounce computations for
  smith microfacet BSDFs}.
\newblock \bibinfo{journal}{\emph{ACM Transactions on Graphics (TOG)}}
  \bibinfo{volume}{41}, \bibinfo{number}{4} (\bibinfo{year}{2022}),
  \bibinfo{pages}{1--14}.
\newblock


\bibitem[Wright(2019)]%
        {Ranger}
\bibfield{author}{\bibinfo{person}{Less Wright}.}
  \bibinfo{year}{2019}\natexlab{}.
\newblock \bibinfo{title}{Ranger - a synergistic optimizer.}
\newblock
  \bibinfo{howpublished}{\url{https://github.com/lessw2020/Ranger-Deep-Learning-Optimizer}}.
\newblock


\bibitem[Xie and Hanrahan(2018)]%
        {Feng2018multiV}
\bibfield{author}{\bibinfo{person}{Feng Xie} {and} \bibinfo{person}{Pat
  Hanrahan}.} \bibinfo{year}{2018}\natexlab{}.
\newblock \showarticletitle{Multiple Scattering from Distributions of Specular
  V-Grooves}.
\newblock \bibinfo{journal}{\emph{ACM Trans. Graph.}} \bibinfo{volume}{37},
  \bibinfo{number}{6}, Article \bibinfo{articleno}{276} (\bibinfo{year}{2018}),
  \bibinfo{numpages}{14}~pages.
\newblock


\bibitem[Xie et~al\mbox{.}(2019)]%
        {xie2019multiple}
\bibfield{author}{\bibinfo{person}{Feng Xie}, \bibinfo{person}{Anton
  Kaplanyan}, \bibinfo{person}{Warren Hunt}, {and} \bibinfo{person}{Pat
  Hanrahan}.} \bibinfo{year}{2019}\natexlab{}.
\newblock \showarticletitle{Multiple scattering using machine learning}.
\newblock In \bibinfo{booktitle}{\emph{ACM SIGGRAPH 2019 Talks}}.
  \bibinfo{pages}{1--2}.
\newblock


\bibitem[Zheng et~al\mbox{.}(2021)]%
        {zheng2021compact}
\bibfield{author}{\bibinfo{person}{Chuankun Zheng}, \bibinfo{person}{Ruzhang
  Zheng}, \bibinfo{person}{Rui Wang}, \bibinfo{person}{Shuang Zhao}, {and}
  \bibinfo{person}{Hujun Bao}.} \bibinfo{year}{2021}\natexlab{}.
\newblock \showarticletitle{A Compact Representation of Measured BRDFs Using
  Neural Processes}.
\newblock \bibinfo{journal}{\emph{ACM Transactions on Graphics (TOG)}}
  \bibinfo{volume}{41}, \bibinfo{number}{2} (\bibinfo{year}{2021}),
  \bibinfo{pages}{1--15}.
\newblock


\bibitem[Zhu et~al\mbox{.}(2022)]%
        {fur2022aggre}
\bibfield{author}{\bibinfo{person}{Junqiu Zhu}, \bibinfo{person}{Sizhe Zhao},
  \bibinfo{person}{Lu Wang}, \bibinfo{person}{Yanning Xu}, {and}
  \bibinfo{person}{Yan Ling-Qi}.} \bibinfo{year}{2022}\natexlab{}.
\newblock \showarticletitle{Practical Level-of-Detail Aggregation of Fur
  Appearance}.
\newblock \bibinfo{journal}{\emph{ACM Transactions on Graphics (Proceedings of
  SIGGRAPH 2022)}} \bibinfo{volume}{41}, \bibinfo{number}{4}
  (\bibinfo{year}{2022}).
\newblock


\end{thebibliography}

\end{document}